\documentclass[fleqn,usenatbib]{mnras}

\usepackage[T1]{fontenc}

\DeclareRobustCommand{\VAN}[3]{#2}
\let\VANthebibliography\thebibliography
\def\thebibliography{\DeclareRobustCommand{\VAN}[3]{##3}\VANthebibliography}

\usepackage{graphicx}	
\usepackage{amsmath}	
\usepackage{amssymb}	
\usepackage{comment}
\usepackage[normalem]{ulem}
\usepackage{newtxtext,newtxmath}
\usepackage[export]{adjustbox}

\newcommand{\Msun}{\ensuremath{\, \rm M_\odot}}

\newcommand{\cmc}{\ensuremath{\,\rm cm^{-3}}}

\newcommand{\pc}{\ensuremath{\,\rm pc}}

\newcommand{\cs}{\ensuremath{c_{\rm s}}}
\newcommand{\ergl}{\ensuremath{\,\rm erg\, s^{-1}}}

\newcommand{\object}[1]{#1}

\newcommand{\amb}{\ensuremath{_{\rm c}}}
\newcommand{\jet}{_{\rm j}}
\newcommand{\hbeta}{\ensuremath{\bar{\beta}_{\rm h}}}
\newcommand{\ra}{\ensuremath{r_{\rm a}}}

\newcommand{\A}{{\cal A}}
\newcommand{\Mach}{{\cal M}}
\newcommand{\pardir}[2]{\ensuremath{\frac{\uppartial #2}{\uppartial #1} }}
\newcommand{\ppardir}[2]{\ensuremath{\frac{\uppartial }{\uppartial #1} \left( #2\right)}}

\renewcommand{\i}{\ensuremath{{\rm i}}}
\newcommand{\diff}{\ensuremath{{\rm d}}}
\renewcommand{\lg}[1]{\ensuremath{\rm log_{10} #1}}
\renewcommand{\deg}{\ensuremath{^\circ}}
\usepackage{xcolor}

\newcommand{\ddr}{\ensuremath{\left\langle \frac{\Delta R}{R}\right\rangle}}

\newcommand{\zrec}{\ensuremath{z_{*}}}

\newcommand{\modelid}[1]{{\it #1}}

\defcitealias{MAP17}{MAP}

\title[]{Interface instabilities in hydrodynamic relativistic jets}
\author[Abolmasov \& Bromberg]{
Pavel Abolmasov\thanks{E-mail: pavel.abolmasov@gmail.com}, 
Omer Bromberg
\\
The Raymond and Beverly Sackler School of Physics and Astronomy, Tel Aviv University, Tel Aviv 69978, Israel
}

\date{Accepted XXX. Received YYY; in original form ZZZ}

\pubyear{2022}

\begin{document}
\label{firstpage}
\pagerange{\pageref{firstpage}--\pageref{lastpage}}
\maketitle

\begin{abstract}
Both the dynamics and the observational properties of relativistic jets are determined by their interaction with the ambient medium. 
A crucial role is played by the contact discontinuity at the jet boundary, which in the presence of jet collimation may become subject to Rayleigh-Taylor instability (RTI) and Richtmyer-Meshkov instability (RMI). 
Here, we study the evolution of these instabilities in non-magnetized relativistic jets using special relativistic three-dimensional hydrodynamic simulations.
We show that the growth of initial perturbations is consistent with relativistic RTI operating in the jet collimation region. 
The contribution of RMI becomes important downstream from the collimation shock in agreement with the theoretical expectations. 
Both instabilities reach non-linear scales above the shock convergence point and trigger strong turbulence, mixing the jet  the with ambient matter.
We devise an analytic solution for the mixing rate and show that it is sensitive to the external density gradients. 
Our results may be applied to different types of astrophysical objects. 
In particular, different contribution of interface instabilities is a natural explanation for the observed dichotomy between FR-I and FR-II radiogalaxies.
The rapid slow-down in the jet of \object{M87} is consistent with baryon entrainment from the circumnuclear matter with the observed density distribution.
In microquasars, baryon loading triggered by interface instabilities is a probable reason for the low observed Lorentz factors.
We show that the observed variability in gamma-ray bursts cannot come from mixing driven by interface instabilities and likely originates from the engine, suggesting the presence of magnetic fields in the jet. 
\end{abstract}




\section{Introduction}

Relativistic jets are a ubiquitous phenomenon primarily associated with mass accretion onto rapidly rotating black holes (BHs). 
They may be powered by stellar-mass BHs, as in the case of microquasars (see review by \citealt{corbel11}) or gamma-ray bursts (see \citealt{piran_review} for a review), or by supermassive BHs, as in tidal disruption events (TDEs, see for example \citealt{colle_review}) or active galactic nuclei (AGNs, \citealt{blandford_review}).
The high observed power and Lorentz factors of relativistic jets suggest an electromagnetic mechanism tapping the rotational energy of the BH to the relativistic outflow, most likely by magnetic fields threading the BH horizon, as in the Blandford-Znajek mechanism \citep[BZ]{BZ}.
If the BH accretes through an accretion disc, launching relativistic jets may be accompanied by a similar Blandford-Payne mechanism \citep[BP]{BP} draining energy and angular momentum from the disc. 
While the BZ mechanism launches Poynting-flux-dominated jets, BP-generated outflows that are baryon-rich and only mildly relativistic. 

Launching jets via BZ mechanism produces a strongly magnetised, essentially force-free outflow, where most of the power released by the central engine propagates in the form of Poynting flux.
However, in most cases when observational data allow analysis of the jet magnetisation, kinetic energy flux seems to dominate over Poynting flux. 
The most robust example are the spectral energy distributions of blazars, where large Compton dominances suggest that relativistic electrons carry more energy than electromagnetic fields \citep{celotti08}.  
Details of jet morphology in radiogalaxies are also interpreted as an indication of predominately kinetic power \citep[e.g.][]{sikora05}. 
Apparently, the energy of the electromagnetic fields is efficiently converted to heat and bulk motion.
Where and how the conversion occurs is still an open question. 

There are numerous ways to dissipate electromagnetic energy in jets. 
If the initial jet magnetisation is high, confinement by the ambient medium incites various types of instabilities like kink instability \citep[e.g.][]{Begelman98,Lyubarskii99,appl00, BT16}, which can dissipate a sizeable fraction of the jet magnetic energy through deformation of the magnetic field lines and subsequent reconnection.
The efficiency of magnetic reconnection depends on the structure and topology of the magnetic field.
For large-scale, ordered field, the total amount of dissipation is limited by flux conservation.
For this reason, polarity changes in the accreted magnetic loops are crucial for efficient dissipation, both in the vicinity of the BH and in the jet itself \citep[e.g.][]{drenkhahn02, Parfrey15, mahlmann20, chashkina21}.
In this case the jet dynamics can become dominated by kinetic rather than magnetic forces, making it suitable for hydrodynamic treatment. 
Alternatively, jets may become effectively hydrodynamic due to additional processes that directly increase their kinetic power.
In particular, in GRB jets, high accretion rates may overload the field lines threading the BH and render the BZ process inefficient \citep{GL13, Gottlieb22}. 
In this case, the jets are expected to be powered by energy sources independent of the magnetic fields and the rotation of the black hole, such as neutrino annihilation \citep{LG13}, which can lead to launching of relativistic hydrodynamic jets.

Hydrodynamic jets formed by the processes discussed above are initially
relativistically hot and quickly become supersonic. 
In this case most of the jet power is stored in kinetic bulk motion and may be dissipated only through interaction with the ambient medium.   
The propagation of a jet through the medium leads to the formation of a system of two shock waves, a forward bow shock in the ambient gas and a reverse shock in the jet, with a contact discontinuity between them.
This compact interaction region is usually called the jet head.
Shocked ambient and jet matter stream sideways from the head and form 
an overpressured cocoon that surrounds the jet and affects its dynamics \citep{BC89}.
At the base of the jet, the jet pressure is much larger than the pressure within the cocoon. 
Expansion and acceleration decrease the pressure inside the jet until it becomes comparable to or smaller than the cocoon pressure.
As both pressures become comparable, the cocoon begins to compress the jet and slow down its sideways expansion, which leads to the collimation of the jet.
The collimation typically occurs when the jet is supersonic, and thus is accompanied by at least one oblique collimation shock \citep[e.g.][]{KF97,Bromberg07, bromberg11}.
The initial collimation is followed by a subsequent expansion of the jet, and potentially a series of collimation episodes (see, for example, \citealt{mizuno15}), analogous to `Mach diamonds' observed in rocket exhaust plumes.

The jet-cocoon boundary is essentially a contact discontinuity, where material on both sides is at an approximate pressure balance, while the rest-mass density changes considerably between the dilute jet material and the relatively dense cocoon. 
Strong gradients of density, entropy, and velocity make the jet boundary a perfect site for interface instabilities.
A fully developed instability allows the jet to exchange matter, momentum, and energy with the cocoon, thus dissipating its kinetic energy. 
This leads to heating of both the cocoon and the jet, and also affects the composition of the jet through entrainment of baryon-rich matter.

Classic BZ models predict relativistic jets to have a light composition with very small fraction of energy carried by baryons.
Thus any observational evidence for a significant baryon content is interesting from the point of view of their origin.
An important indirect argument for baryon loading in AGN jets is the growing evidence for the association of ultra-high-energy cosmic rays (UHECR) and high-energy neutrinos with the population of AGNs \citep{rodrigues21}. 
Also, at least in one case \citep{nu_TXS}, a gamma-ray flare from a blazar was associated with a neutrino flare detected by {\it IceCube}.
At the same time, there are strong arguments against large baryon content in blazar jets based on total available power budget and radiation models (see \citealt{zdziarski22} and references therein).
It is unclear if the proton content of the jets is inherited from the central engine or drawn from the ambient medium. 
In the latter case, surface instabilities are a crucial factor for the evolution of the baryon content, but also an important factor slowing down and disrupting the jets.

Different kinds of hydrodynamic instabilities are expected to be important at the jet boundary.
Existence of a tangential velocity jump allows for the growth of Kelvin-Helmholtz instability (KHI). 
If, however, the jet remains relativistic at its boundary, the velocity jump is supersonic, and KHI is suppressed \citep{bodo04}. 
The collimation of the jet by the cocoon pressure creates an acceleration field on the boundary directed towards the jet axis, which leads to an effective gravity in the opposite direction. 
In the presence of a density jump, the boundary may be subject to Rayleigh-Taylor instability (RTI). 
Last, the collimation shock is expected to converge towards the axis. 
For an axisymmetric jet, the shock meets the axis at a point $z=\zrec$ we call the \emph{convergence point} (also known as reconfinement or recollimation point).
Above this point, the shock extends further outward (this can be viewed as a reflection from the jet axis), until it crosses the contact discontinuity, driving the Richtmyer-Meshkov instability (RMI). 
The importance of RMI in collimated relativistic jets was stressed by \citet{MM13}. 

Linear stability analysis of relativistic RTI done by \citet[hereafter \citetalias{MAP17}]{MAP17} has shown that the stability of the contact discontinuity to RTI is determined not only by the density contrast but also by the Lorentz factor of the jet and its internal energy. 
As a result, even though the jet is much lighter than the ambient medium (in terms of rest-mass density), it can behave as a `relativistically heavy' fluid that sits on top of a `lighter' fluid and drives RTI fingers into the cocoon.
This kind of behaviour was later confirmed by numerical simulations by, e.g., \citet{MM19} and \citet{Gottlieb21}.

The nature of the relativistic RTI in jets is essentially three-dimensional, as it involves the jet propagation direction, the direction of the effective gravity, and the direction of the wave vector of the unstable mode, where the three can not be co-planar. 
Hence, a self-consistent numerical study should be three-dimensional, and resolve both the growing perturbations and the global structure of the collimation region. 
Global 3D simulations by \citet{MM19} and \citet{Gottlieb21} considered the propagation of a relativistic jet in a static medium. 
In such simulations the cocoon appears naturally, and its interaction with the jet was shown to generate structures on the jet boundary that are expected to appear as an outcome of RTI. 
These simulations have shown that RTI leads to the mixing of jet and cocoon material, which reduces the jet specific enthalpy and limits its ability to accelerate to high Lorentz factors. 
The growth of the initial perturbations was strongly suppressed in cases when linear stability analysis predicts the contact discontinuity to be stable against RTI. 

In the numerical simulations discussed above, the initial perturbations normally grow with time and downstream along the jet. 
This growth is probably related to RTI, but even for the most detailed simulations there are no quantitative estimates for the growth rates. 
Besides, the effects of jet boundary stability in such simulations are mixed with the effects of jet head propagation and time evolution of the cocoon. 
An important test for the RTI hypothesis would be a steady-state headless jet model, where the growth of the perturbations may be studied as a function of the coordinate along the jet. 
Such a solution is impossible in a static medium, where the cocoon proceeds expanding, the pressure inside it drops, and the collimation properties of the jet change with time.
Instead, we consider a jet propagating in a steady sub-relativistic flow.
In this case, as we will show below, the system can reach a steady state, and the time evolution of the RTI may be studied as the spatial structure of the jet.

There are physical reasons to expect the external medium to be moving. 
Many AGNs are known to launch mildly relativistic, apparently non-collimated, ultra-fast outflows (UFOs, see \citealt{tombesi10}). 
Their origin can naturally be related to BP or to some other types of winds from the inner portions of accretion discs \citep{tombesi_review, fukumura}.
Different kind of winds, colder and denser, are observed in broad-absorption-line (BAL) quasars \citep{murray95}.
Any kind of a wind launched by the inner accretion disc creates an environment that the jet is bound to interact with. 
Such winds may be responsible for jet collimation \citep{Levinson00,Bromberg07,globus16} directly or through their interaction with the cocoon. 
Here, we do not specify the nature of the fluid surrounding the jet, but assume it is moving non-relativistically in the same direction as the jet. 
We will refer to such a concept as \emph{dynamic cocoon}.
The main goal of the paper is to preform a self-consistent numerical study of the steady-state structure of a hydrodynamic jet interacting with a dynamic cocoon.
We focus on the instabilities that develop on the interface between the jet and the cocoon, obtaining the growth rates and comparing them with analytic estimations for different types of instability.
We apply our results to different types of astrophysical jets evolving in different environments.

The paper is organised as follows. 
In Section~\ref{sec:RTI}, we outline the physical picture of RTI evolution in a steady-state jet.
In Section~\ref{sec:sim}, we describe our numerical setup, and present the simulation results in Section~\ref{sec:res}. 
In Section~\ref{sec:disc}, we apply our results to AGNs, X-ray binaries, and GRBs, and make conclusions in Section~\ref{sec:conc}.

\section{RTI in relativistic flow}\label{sec:RTI}

\subsection{Stability and linear growth}\label{sec:RTI:linear}

Classic RTI (see, e.g., \citealt{sharp_RTI} for a review) considers a uniform gravity field $\mathbf{g}$ and a contact discontinuity surface orthogonal to the direction of the gravity vector. 
All the motions are assumed to be subsonic and in approximate pressure equilibrium, while the growth of the perturbations is caused by the variable contribution from buoyancy force. 
The contact discontinuity at the jet boundary fits these conditions, with an important difference of having a supersonic velocity jump in the tangential velocity component. 
The local stability problem of the jet boundary may be solved in the assumption of negligibly small gradients in the direction of the flow, where the role of relativistic motions is reduced to changing the inertial properties of the flow according to its Lorentz factor, $\gamma$.
This allows to treat the problem as two-dimensional, as it was done by \citetalias{MAP17}.

Consider the stability of a cylindrical contact discontinuity in a flow moving along the $z$ axis. 
The wave vector of the unstable mode points along the azimuthal direction, while the perturbations are radial. 
A linear stability analysis of such a system was preformed by \citetalias{MAP17} in the cocoon frame.
Their analysis allowed to express the perturbation growth rate, $\sigma$, in a form typical for Rayleigh-Taylor instability,
\begin{equation}\label{E:RTI:sigma}
    \sigma = \sqrt{\A g k},
\end{equation}
where $g$ is gravity or inertial force measured in the cocoon reference frame, $k$ is the wave number of the perturbation, and $\A$ is the relativistic Atwood number. 
Note that due to the boundary conditions for the perturbations, $k$ should be a real and positive number, while the 
gravity $g$ is positive if directed outward, away from the jet axis.
The relativistic Atwood number can be expressed, according to \citetalias{MAP17}, as
\begin{equation}\label{E:RTI:A}
    \displaystyle \A = \frac{\left(\gamma^2h^\prime \rho\right)_{\rm j}-\left(\gamma^2h^\prime \rho\right)\amb}{\left(\gamma^2 h \rho\right)_{\rm j}+\left(\gamma^2h \rho\right)\amb},
\end{equation}
where indices `j' and `c'  correspond, respectively, to the jet and the dynamic cocoon. 
Here, $\rho$ is the density, $\gamma$ is the Lorentz factor, 
\begin{equation}\label{E:RTI:h}
    h = 1 + \frac{\Gamma}{\Gamma-1} \frac{P}{\rho}
\end{equation}
is the specific enthalpy, and 
\begin{equation}\label{E:RTI:hprime}
    h^\prime = 1 + \frac{\Gamma^2}{\Gamma-1} \frac{P}{\rho},
\end{equation}
where $P$ is pressure, assumed to follow the adiabatic law $P \propto \rho^\Gamma$.
The sign in Eq.~(\ref{E:RTI:sigma}) is chosen in a way that $\sigma$ is real (and the flow is unstable) when $\A >0$.
This is the \emph{heavy jet} case, when the inertia of the jet exceeds that of the ambient medium. 
The opposite case of a \emph{light jet} ($\A < 0$) is expected to be stable to RTI in the recollimation zone.
The combinations $\gamma^2h \rho$ and $\gamma^2h^\prime \rho$ replace the densities in the classic Newtonian RTI. 
Hereafter, we will refer to $\gamma^2h\rho$ as \emph{relativistic inertia}.

Following the reasoning of \citetalias{MAP17}, we consider motions in the cocoon reference frame.
If the cocoon moves at a dimensionless velocity $\beta_{\rm c}$, the local effective gravity is estimated as minus the local radial acceleration of the flow
\begin{equation}
    g = - \frac{\diff^2 R_{\rm d} }{\diff \tau\amb^2},
\end{equation}
where $\tau_{\rm c}$ is proper time in the cocoon frame, and $R_{\rm d} = R_{\rm d}(z)$ is the radial coordinate of the contact discontinuity. 
Streamlines do not cross the discontinuity, which allows to relate the Largangian acceleration to the shape of the surface. 
If the flow is steady, its velocity in the $z$ direction is constant, and the radial velocity is non-relativistic, one may replace $\tau\amb = z/\beta\amb \gamma\amb$ (where $\gamma_{\rm c}$ is the Lorentz factor of the cocoon), implying
\begin{equation}\label{E:RTI:g}
    g = -\gamma_{\rm c}^2\beta_{\rm c}^2 c^2\frac{\diff^2R_{\rm d}}{\diff z^2},
\end{equation}
assuming $\beta_{\rm c}$ is constant.
The effective gravity may be viewed as an inertial force related to the radial acceleration of the flow, or, alternatively, as a centrifugal force.
The latter approach allows to consider relativistic RTI in the jet collimation regions as a centrifugally driven instability, as it was done by \citet{GK18}. 
The second derivative in Eq.~(\ref{E:RTI:g}) then may be viewed as $R_{\rm d}$ over the curvature radius of the arc along which the flow is moving. 

The surface of the jet is perturbed in the direction perpendicular to both the jet direction ($z$ axis) and to the acceleration vector. 
Thus, the direction of the wave vector is along the azimuthal coordinate, and the wavenumber may be expressed as
\begin{equation}
    k = \frac{m}{R},
\end{equation}
where $m$ is a natural number. 
Though the value of $m$ does not affect the stability itself, it affects the growth rate through the value of $\sigma$, thus allowing to check the RTI model for the evolution of the jet. 

In the steady-state case, $\sigma$ describes spatial rather than temporal variations of the perturbation amplitude. 
When $\sigma^2 > 0$, small-amplitude perturbations grow with $z$ as
\begin{equation}\label{E:RTI:dR}
    \eta = \frac{\diff }{\diff z} \ln \Delta R = \frac{\sigma}{\beta_{\rm c}} = \sqrt{- {\cal A} m \frac{1}{R_{\rm d}} \frac{\diff^2 R_{\rm d}}{\diff z^2}}. 
\end{equation} 
The result does not depend on $\beta_{\rm c}$, and should be valid also in the limit $\beta_{\rm c} \to 0$. 

\subsection{The dominant perturbation mode in the non-linear regime}

The growth rate estimate in the previous subsection considers a single azimuthal mode.
In our simulations (see Section~\ref{sec:sim}) we will also restrict the initial perturbation to a single mode that evolves from a relative amplitude of $\sim 0.01$ to a non-linear scale.
In reality, the perturbations will contain a set of all the possible modes with various amplitudes. 
The amplification conditions for different modes are, however, unequal. 
The growth increment scales with the wavenumber as $\sigma \propto \sqrt{k}$, which makes shorter waves grow faster until they reach saturation. 
For each azimuthal mode, the saturation occurs when the amplitude of the radial variations $\Delta R_{\rm d}$ exceeds the mode wavelength.
This sets the maximal azimuthal number for a particular amplitude $m_{\rm max} \sim 2\uppi \frac{R_{\rm d}}{\Delta R_{\rm d}}$. 
For the parameters used in our simulations (see later Section~\ref{sec:sim}), $m_{\rm max} \simeq 100$.
The detailed evolution depends on the initial perturbation spectrum. 
As the azimuthal wavenumbers vary considerably (by about two orders of magnitude), one would expect a self-similar solution to exist, with the dominating mode wave-number decreasing as smaller and smaller $m$ reach saturation. 
For a stratified static fluid in a gravity field, such a self-similar solution with the perturbation amplitude growing with time as $\propto t^2$ was proposed by \citet{Fermi53}. 
The wavenumber of the dominating mode in this regime decreases as $k\propto t^{-2}$.

Detailed analysis of the multi-mode perturbation growth in
an axisymmetric collimated jet is given in Appendix~\ref{sec:app:fermi}. 
If the range of available modes is large enough, one might expect a self-similar asymptotic with a dominant mode number scaling with distance as $m\propto z^{-3}$ to work in the collimation region. 
The most successful mode downstream the collimation region would have the number of $\displaystyle m \sim \frac{2\ln^2\delta}{\uppi^2\A}$ (where $\displaystyle \delta = \frac{\Delta R_0}{2\uppi R_0}$ is a normalized initial perturbation amplitude) or $m_{\rm max}$, whichever is smaller (see App. ~\ref{sec:app:fermi}) . 
Substituting the wave number of the most successful mode in Eq.~(\ref{E:RTI:dR}) gives a characteristic optimal perturbation growth rate
\begin{equation}\label{E:sigma:opt}
    \left. \frac{\diff }{\diff z} \ln \Delta R\right|_{\rm opt} = \frac{|\ln \delta|}{\uppi} \sqrt{-\frac{1}{R_{\rm d}}\frac{\diff^2 R_{\rm d}}{\diff z^2}} \sim \frac{|\ln \delta|}{\uppi} \frac{1}{z_{\rm c}},
\end{equation} 
where $z_{\rm c}$ is the characteristic collimation length, meaning that \emph{the spatial scale for the linear growth of RTI in realistic RTI-unstable jets should be of the order $z_{\rm c}$}. 
Beyond this distance ($z\gtrsim z_{\rm c}$), perturbations are strongly non-linear in the sense that $\Delta R \sim R_{\rm d}$.

\subsection{External mass entrainment in the non-linear regime}\label{sec:RTI:NL}

Interface instability leads to jet mixing with the environment, which is important both for the dynamics of the jet (energy is dissipated and lost, while mass is gained) and for its observational properties. 
Our simulations cover the linear stage of RTI and extend into the non-linear regime.
The structures formed by RTI have velocities limited by the speed of sound, as supersonic motions are dumped by shock waves. 
If the sound speeds are different on the two sides of the unstable interface, the lower value is the most important. 

The fluxes of mass and energy depend on the definition of the jet surface. 
This definition may be more or less inclusive, as mixing creates a growing region intermediate in its properties between the jet and the cocoon.
Let us define the boundary in a way that energy does not leave the jet, but there is mass entrainment. 
For a jet with a cylindrical radius $R_{\rm j}$, the mass flux into the jet through a unit surface area on the jet boundary is
\begin{equation}\label{eq:dMdot_dz}
    \frac{1}{2\uppi R_{\rm j}}\frac{\diff \dot{M}}{\diff z} \simeq \left(\rho\amb - \rho_{\rm j}\right) \cs, 
\end{equation}
where $\cs$ is the speed of sound in the ambient medium (presumably, lower than the speed of sound in the jet). 
The mixing length may be defined as the characteristic length scale over which the mass flow in the jet $\dot{M_j}=\rho_j\uppi R_j^2 u^z$ doubles.
Substituting this in Eq.~(\ref{eq:dMdot_dz}), we get
\begin{equation}\label{E:zmix}
    z_{\rm mix} \sim  \frac{\rho_{\rm j}}{2\left(\rho\amb-\rho_{\rm j}\right)} \frac{u^z}{\cs} R_{\rm j} \simeq \frac{\rho_{\rm j}}{2\rho\amb} \frac{u^z}{\cs} R_{\rm j},
\end{equation}
where the original rest-mass density of the jet was assumed to be negligibly small.

The actual mass gain rate depends on the shape of the jet. 
In particular, for $R_{\rm j} = \theta_{\rm j} z$, where $\theta_{\rm j}= $ const is the jet opening angle, 
\begin{equation}\label{E:disc:mdot}
    \dot{M} \simeq \dot{M}_0 + \uppi \theta_{\rm j} \rho\amb \cs z^2,
\end{equation}
with a characteristic mixing length of
\begin{equation}\label{E:disc:zm}
    z_{\rm mix} \sim \sqrt{\frac{L\jet}{\uppi \theta_{\rm j} \gamma_\infty \rho\amb\cs}} \simeq \sqrt{\frac{\cs}{\theta_{\rm j}\gamma_\infty}} \zrec,
\end{equation}
where $L\jet$ is thew total power of the jet, $\gamma_\infty = h\gamma$ is relativistic Bernoulli parameter, conserved along a streamline in the absence of mixing, and $\zrec$ is the position of the collimation shock convergence point.
For $\zrec$, we used the estimate from \citet{bromberg11} 
\begin{equation}\label{E:disc:zrec}
    \zrec \sim z_{\rm c} \simeq \sqrt{\frac{\beta\jet L\jet}{\uppi c P\amb}}.
\end{equation}
For a fully relativistic jet with $\beta\jet \simeq 1$, $\gamma_\infty \theta_{\rm j} \sim 1$, and $\cs \lesssim 1$, $z_{\rm mix}$ is about or smaller than the collimation length. 
Note that this scale becomes relevant only after the interface instabilities enter the non-linear regime.
Transition to the non-linear regime occurs at 
$z\gtrsim z_{\rm c}$ and may be related either to RTI or to RMI.
RMI is excited after the passage of the reflected collimation shock through the discontinuity, and its growth rate is calculated in Appendix~\ref{sec:app:RM}. 

Mass entrainment leads to the formation of a two-component jet consisting of a relativistic core, which retains its initially high Lorentz factor, and a slower sheath that holds the bulk of the entrained material. 
As the jet evolves it tends to preserve its total power, while the mass-flow rate $\dot{M}(z)$ increases due to entrainment, hence the average Bernoulli parameter in the jet, defined as
\begin{equation}\label{eq:gamma_inf_avv}
<\gamma_\infty>=\frac{L\jet}{\dot{M}(z)},
\end{equation}
decreases. 
The jet core, on the other hand, loses energy to the sheath while hardly gaining any mass\footnote{This may equally be considered the energy loss from the relativistic jet, if we relax the ``maximally inclusive'' jet definition used above.}.
Let us define the core as the region where the Bernoulli parameter exceeds some cut-off value $\gamma_{\infty}^{\rm c}$.
If the fraction of jet power within the core decreases proportionally to its mass flow fraction, the power decreases with $z$ as 
\begin{equation}
    L(\gamma_\infty > \gamma_\infty^{\rm c}) \simeq L_{0,{\rm c}}\frac{\dot{M}_0}{\dot{M}(z)},
\end{equation}
where $\dot{M}_{0}$ and $L_{0,{\rm c}}$ are the jet mass flow and core power at the injection point.
For a cylindrical jet ($R\jet = $const), $\dot{M}$ grows linearly with $z$,
\begin{equation}\label{eq:dot_M_lin}
    \dot{M}(z) \simeq \dot{M}_0 + 2\uppi R\jet \rho\amb \cs z,
\end{equation}
and the core power decreases as
\begin{equation}\label{E:NL:ploss}
    L(\gamma_\infty > \gamma_\infty^{\rm c}) \simeq \frac{L_{0,{\rm c}}}{1+ z/\Delta z},
\end{equation}
where $\Delta z$ is defined as the typical scale over which the jet mass flow doubles. 
As we will see in Section~\ref{sec:res:hg}, such a fractional dependence provides a good approximation to the simulation results. 

Because the expected values of $z_{\rm mix}$ are generally smaller than $z_*$ (see Eq.~\ref{E:zmix}), as soon as the fluid interface instabilities fully develop (which happens on a $\sim z_{\rm c}$ scale), a hydrodynamic jet in a uniform medium rapidly decays and decelerates.
After several collimation lengths, one might expect most of the jet power to be spent on exciting turbulence via interface instabilities. 

\section{Numerical simulations}\label{sec:sim}

The simulations were run using the GRMHD code {\sc Athena++}\footnote{The code is freely available at \url{https://github.com/PrincetonUniversity/athena}.} \citep{athena, athenapp} in its special-relativistic hydrodynamic mode (i.e. no gravity and no magnetic fields). 
We use a cylindrical grid with a resolution of $384\times384\times1536$ cells in the $R$, $\varphi$ and $z$ directions, respectively. 
Distances and time are measured in scalable code units, assuming the speed of light $c=1$.
The units for density and pressure are also arbitrary and are normalized in the same way.
In the scalable code units, the grid covers the range from $R_{\rm in}=1.5$ to $R_{\rm out}=150$ in the radial direction, where cells are spaced logarithmically and maintain a constant ratio of $\Delta R/ R \simeq 0.03$.
In the azimuthal direction the cells are evenly spaced, giving a constant aspect ratio of $R\Delta\varphi/\Delta R\simeq 0.5$, while in the $z$ direction we use a uniform grid between of $z_{\rm min}=0$ to $z_{\rm max}=200$ with $\Delta z \simeq 0.13$. 
The resulting aspect ratio in the meridional plane $\Delta z/\Delta R$ changes with radius, and is close to unity at $R=5$.
In one of the models (\modelid{U3e}, see Table~\ref{tab:mod}), the mesh is extended to $z_{\rm max}= 300$, with the number of cells increased to $N_z=1728$, resulting in slightly longer cells with $\Delta z \simeq 0.17$. We use reflective boundary conditions in the radial direction, periodic in the azimuthal direction, outflow at $z_{\rm max}$ (upper boundary), and inflow at $z_{\rm min}$ (lower boundary).
The boundary conditions at $z=z_{\rm min}$ play the role of initial conditions and determine the properties of both the jet and the cocoon in the steady state.
Reflective boundary condition at $R_{\rm in}$ allows to avoid the coordinate singularity on the axis ($R_{\rm in}\gg \Delta R$) and does not interfere with the surface of the jet and its stability considered in this paper. 
However, global oscillation modes involving the bulk of the jet may be affected, as the core of the jet can not move sideways freely in the presence of a reflecting core around its axis.
 
The computational domain is initially filled with a medium of uniform pressure $P\amb=0.1$ and mass density $\rho\amb=1$ or $100$ for the `heavy-jet' and `light-jet' models, respectively. 
To allow the system to reach a steady state, the medium is assigned with a uniform vertical 4-velocity $u\amb^z=0.1$.
The list of models used in this work is given in Table~\ref{tab:mod}, where the naming code consists of the letter `U' for unstable (`heavy-jet') or `S' for stable (`light-jet') models, according to the expectations from the linear RTI theory (\citetalias{MAP17}), the azimuthal wave number $m$ of the initial perturbation (3 or 27), and potentially an additional letter specifying the modifications of the model (rotation, grid extension, or the structure of the transition region).

At the onset of the simulation, we start injecting matter from the entire lower boundary, creating the jet and replenishing the ambient medium moving upwards.
Jet material is injected at $R<R_{\rm j,0}=5$ with a rest-mass density $\rho\jet = 10^{-2}$, pressure $P\jet = 0.1$, and $u\jet^z=5$. 
At $R>R_{\rm j,0}$ we inject the dynamic cocoon with the properties matching the initial conditions.
We use a polytropic equation of state $P\propto \rho^\Gamma$ with a constant adiabatic index $\Gamma=4/3$, resulting in a specific enthalpy of $h\jet = 1+4P/\rho \simeq 41$ for the jet material and $h\amb=1.004$ or $1.4$ for the cocoon in stable (`light jet') and unstable (`heavy jet') models, respectively.

As the jet and the dynamic cocoon are in pressure balance at injection, collimation requires initial expansion of the jet. 
We assume that the jet is initially expanding homologously with a radial 4-velocity $u\jet^R  = \alpha u^z R / R\jet$, where $\alpha = 0.25$. 
In the rotating jet models \modelid{U27r} and \modelid{S27r}, we add a tangential velocity component inside the jet with a constant value $\Omega = u^\varphi / u^t = 0.1$ (see Table~\ref{tab:mod}). 
To avoid numerical errors due to large gradients at the jet boundary and to be able to control the initial perturbations, we treat the jet-cocoon boundary as a smooth transition. 
For the transition, we use a smoothing function of the form
\begin{equation}
    s(x) = \frac{\tanh x + 1}{2},
\end{equation}
where $x$ is the radial distance normalized by the thickness of the transition region.
As we are interested in controlling the RTI growth rate, the smoothing is done on the relativistic inertia $A=\gamma^2h\rho$ and modified relativistic Bernoulli parameter 
\begin{equation}
    u_\infty = \sqrt{(h\gamma)^2-1},
\end{equation}
which has an advantage over $\gamma_\infty$ of being sensitive to the properties of sub-relativistic flows.
The same kind of smoothing is used for $\alpha$ and $\Omega$.
As the pressure across the boundary is uniform, these conditions set the radial dependence of the density and velocity.
Details of the smooth transition structure are given in Appendix~\ref{sec:app:boundary}.

\begin{table}\centering
\caption{Parameters of the simulations, from left to right: model ID; ambient density $\rho\amb$; injected perturbation wave number $m$; injected jet angular velocity $\Omega$;  relativistic Atwood number on the jet boundary $\A$; simulation time intervals in code units $t$. 
Models \modelid{U27}, \modelid{S3} and \modelid{S27} do not start from $t=0$ but extend other simulations. 
Simulation \modelid{U27} starts from the last snapshot of simulation \modelid{U3}, while simulations \modelid{S3} and \modelid{S27} extend \modelid{S27s}. 
This is done to obtain faster relaxation into steady states.}\label{tab:mod}
\begin{tabular}{lcccccc}
\hline \hline
ID & $\rho\amb$ & $m$ & $\Omega$ & $\cal A$ & $t$ & comment\\
\hline
U3 & $1$ & $3$ &$0$ & $1.03$ & $0 - 1800$\\
U27 & $1$ & $27$& $0$ & $1.03$ & $1800 - 5000$ \\
S3 & $100$ & $3$ &$0$ & $-0.79$ & $5000 - 6500$  \\
S27 & $100$ & $27$ &$0$ & $-0.79$ & $ 5000 - 6700 $\\
S27s & $100$ & $27$ &$0$ & $-0.79$ & $0 - 5000$ &{\footnotesize step in $\rho$ and $u^z$}\\
U27r & $1$ & $27$& $0.1$ & $1.03$ & $0 - 1800$ \\
S27r & $100$ & $27$& $0.1$ & $-0.79$ & $0 - 3000$ \\
U3e  &  $1$  & $3$ & $0$   & $1.03$  & $0-4000$ & $N_z=1728$ \\
\end{tabular}
\end{table}

To start the RTI, we perturb the jet radius in the injection zone ($z=0$) by adding a sinusoidal modulation function in the azimuthal direction: $R\jet(z=0)\equiv R_{\rm j,0}(1+0.01 \cos(m\varphi))$, where $m$ is the wavenumber of the initial perturbation. 
The wavenumber was chosen to be $3$ or $27$ for different models as shown in Table \ref{tab:mod}. 
Note that using aligned cylindrical coordinate frame allows to control the shape of the perturbations. 
Even in high-resolution simulations using Cartesian grid, perturbations created by the grid act as the seeds for interface instabilities (this is clearly visible, for instance, in \citealt{Gottlieb20b}).

Each simulation is evolved towards a steady state, which can be reached after the jet head exits the computational domain. 
At this point the jet effectively becomes headless, and the entire flow rapidly becomes steady.
The two characteristic time scales for the relaxation of the jet are the crossing time of the cocoon $t\amb \sim z_{\rm max} / \beta\amb \sim 2000$ and the sound-crossing time $\displaystyle t_{\rm s} \sim \sqrt{\frac{\rho\amb}{\Gamma P\amb}} z_{\rm max}$. 
For the parameters used in the simulations, $t_{\rm s}$ ranges between several hundreds to several thousands.
As we show in Section~\ref{sec:res:jet}, the flow evolves to a steady state on time scales comparable or even shorter than $t_{\rm c, \, s}$, allowing us to study the growth of interface instabilities using spatial information only.

\section{Results}\label{sec:res}

\subsection{Jet collimation and expansion}\label{sec:res:jet}

\begin{figure}
\adjincludegraphics[width=1.0\columnwidth, trim = {{0.0\width} {0.0\height} {0.0\width} {0.0\height}}, clip]{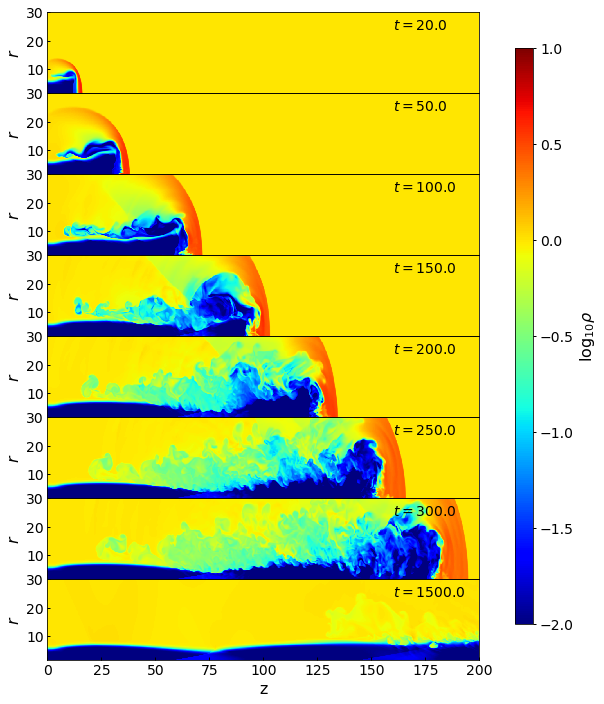}
 \caption{Eight snapshots showing jet propagation through the domain in model \modelid{U3}. 
 The lowest panel ($t=1500$) shows the jet in a steady-state after the head has exited the top boundary. 
 Colour-coded is the logarithm of rest-mass density at $\varphi = 0$.} 
 \label{fig:proplot}
\end{figure}

Fig.~\ref{fig:proplot} demonstrates the evolution of the system at the beginning of the simulation.
For illustrative purposes the axes are rotated by $90^\circ$, so that $z$ axis is horizontal. 
The plot shows meridional cross-sections of the rest-mass density at $\varphi = 0$.
The jet enters from the left side (at $z=0$) and creates a system of shock waves at its head, which heats the ambient medium creating a cocoon structure around the jet.   
Once the jet head leaves the computational domain at $t \sim 300$ (in model \modelid{U3}), the system rapidly relaxes to a steady-state configuration (shown in the lowest panel of Fig.~\ref{fig:proplot}). 

After reaching a steady state, each model has a well-developed collimation region approximately between $z=0$ and a convergence point ($z=\zrec$), where the streamlines are focused towards the jet axis. 
The position of the convergence point 
varies in most models between $30$ and $60$ (see Table~\ref{tab:res}), consistent with Eq.~(\ref{E:disc:zrec}) for $L\jet \simeq 1000$.
This value for the $L\jet$ is a very crude estimate, as most of the jet power is concentrated towards its boundary. 
For example, in model \modelid{S27} the power within $R<5\pm 0.5$ is $L\jet\simeq 1200^{+1300}_{-500}$, respectively, leading to an analytically estimated convergence point $\zrec^{\rm an}(\modelid{S27}) \simeq 50^{+40}_{-5}$. In  model \modelid{C27s}, the transition layer has even larger contribution to the total power with $L_{\rm j}\simeq 6000^{+1500}_{-400}$ for $R<5\pm 0.5$. A similar estimate for the location of the convergence point in \modelid{S27s} yields $\zrec^{\rm an}(\modelid{S27s}) \simeq 140^{+15}_{-60}$. Except for \modelid{S27s}, the position of the convergence point is better predicted by the power in the jet core $R\lesssim 4.5$ (see Table~\ref{tab:res}).

\begin{table*}\centering
\caption{ 
Instability growth rates, convergence point positions, and the fraction of jet energy (total power for $\gamma_\infty > 100$) reaching $z=150$, estimated for different models during the period of time $t_{\rm av}$. 
}
\label{tab:res}\small
\begin{tabular}{lccccc}
\hline \hline
ID & $t_{\rm av}$& \multicolumn{2}{c}{$\eta$} &  $\zrec$& $L_{\rm j}(z=150)/L_{\rm j}(z=0)$ \\
   & & measured & predicted  \\
\hline
U3 & 1500-1800 & $0.02829\pm 0.0003$ & $0.0351\pm 0.0002$ & $53.4\pm 0.9$ & 0.9994$\pm$0.0002\\
U27 & 2000-5000 & $0.0766\pm 0.0009$ & $0.1180\pm 0.0006$ & $53.4\pm 0.2$ & 0.485$\pm$0.016\\
S3  & 5900-6500 & $-0.0086\pm 0.0010$ & $0.0060\pm 0.0003$ & $36.5\pm 0.2$ & 1.004$\pm$0.003\\
S27 & 6000-6600 & $-0.036\pm 0.003$ & $0.020\pm 0.001$ & $36.6\pm 0.2$ & 1.007$\pm$0.004\\
S27s & 3800-5000 & $0.027\pm 0.003$ & $0$ & $196\pm 1.5$ & 1.009$\pm$0.004\\
U27r & 1000-1800 & $0.0539\pm 0.0011$ & $0.1151\pm 0.0008$ & $45.5\pm 0.2$ & 0.710$\pm$0.019\\
S27r & 2500-3000 & $-0.026\pm 0.004$ & $0.0286\pm 0.0014$ & $34.2\pm 0.2$ & 0.834$\pm$0.002\\
U3e  & 3000-4000 & $0.0293\pm 0.0004$ & $0.0350\pm 0.0003$ & $53.6\pm 0.3$ & 1.0001$\pm$0.0002\\
\end{tabular}
\end{table*}

\begin{figure*}
\includegraphics[width=0.9\textwidth]{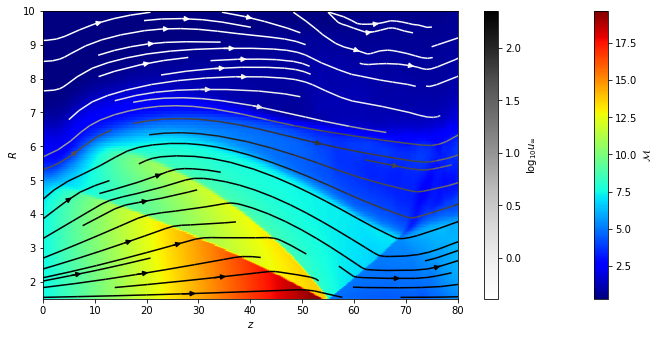}
 \caption{Relativistic Mach number and stream lines for an azimuthally averaged snapshot of the simulation \modelid{U27}. Colours of the lines show the value of $u_\infty$.
 }\label{fig:stream}       
\end{figure*}

In all the `heavy-jet' models, the collimation region has a more intricate structure with two nested collimation shocks instead of one. 
An example for such a structure is shown in Fig.~\ref{fig:stream} for model \modelid{U27}. 
The figure shows a map of the relativistic Mach number \citep{bodo04}
\begin{equation}
    \Mach = \gamma \beta \frac{\sqrt{1-\cs^2}}{\cs},
\end{equation}
where
\begin{equation}
    \cs = \sqrt{\frac{\Gamma P}{h\rho}}   
\end{equation}
is the local sound velocity. 
A structure of two nested oblique shocks converging towards the jet at $\zrec \simeq 53$ is seen in the jet collimation region.
We also show the streamlines coloured according to the modified Bernoulli parameter $u_\infty$. 
Within the jet, the values of $u_\infty$ are well above unity, and the streamlines are dark, while the dynamic cocoon has small values of $u_\infty$, and the streamlines are light grey or white. 
Multiple collimation shocks may occur if the jet flow remains supersonic after crossing the first collimation shock. 
As the shock is oblique, and tangential velocity component changes continuously through the shock front, existence of a compression fan instead of a single conical collimation shock is a probable scenario. 
The most likely reason why such a structure was never obtained in simulations is that our jet initially expands supersonically, with a radial Mach number $\Mach_R = 0.25 \Mach \simeq 2$.
As we are interested in the properties of the contact discontinuity rather than the shock, we will leave further studies of the compression fan structure in relativistic flows to a separate study.  

\begin{figure*}
\includegraphics[width=\textwidth]{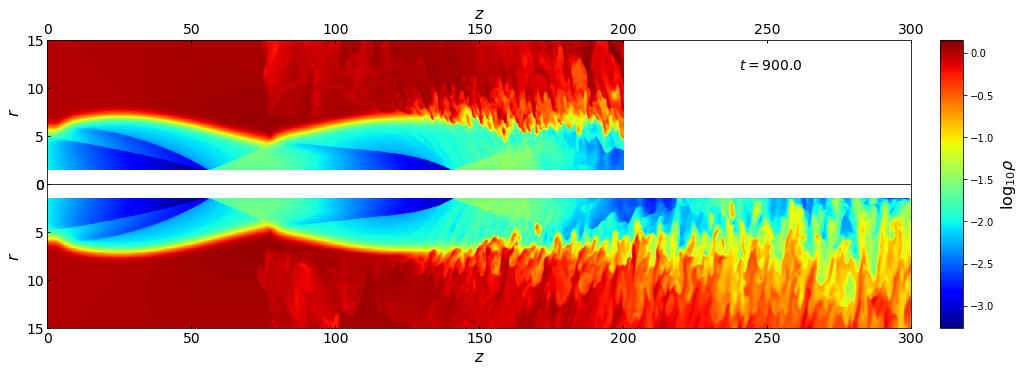}
 \caption{Comparison of the models \modelid{U3} (upper panel) and \modelid{U3e} (lower panel), differing only in the grid parameters ($z_{\rm max}$ and resolution). 
 Meridional cross-section ($\varphi = 0$) of $\log_{10}\rho$  is shown for the same instance of time $t=900$. 
 }\label{fig:complot}       
\end{figure*}

Above the shock convergence point, the jet passes through another weak shock that may be considered as a reflected collimation shock. 
Interaction with this shock triggers RMI at the jet boundary. 
After passing through the reflected shock, 
the jet becomes divergent and enters a second collimation region. 
This well-known picture of cyclic collimations and expansions \citep{mizuno15, MM13} is gradually destroyed by the growing instabilities.  
This is can be seen in Fig.~\ref{fig:complot}, where the jet becomes efficiently mixed after about two collimation cycles. 
The figure also shows the impact of the upper boundary conditions: the two simulation snapshots shown in the plot differ only in the mesh spacing and in their extent ($z_{\rm max} = 200$ vs. $300$), and do not show any noticeable differences within the simulation domain. 

\begin{figure*}
\includegraphics[width=\textwidth]{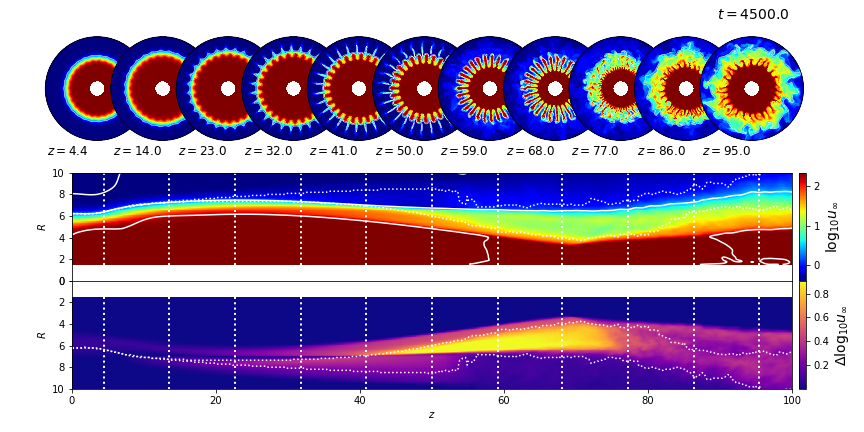}
 \caption{Structure of the flow in the steady-state regime of the simulation \modelid{U27}. All the panels show relativistic Bernoulli parameter $u_\infty$. The upper panels are cross-sections of $z=$const (positions of the cross-sections are shown in the lower plots), and the lower two represent the azimuthal averages and azimuthal root-mean-square variations of $\lg u_\infty$. 
 White solid lines are the contours of (azimuthally averaged) $u^z = 0.1$,  $1$, and $5$. 
 Dotted white lines show the maximal and minimal (over $\varphi$) radial coordinates of the contact discontinuity.
 }\label{fig:steady}       
\end{figure*}

\subsection{Instability growth rates}\label{sec:res:inst}

Fig.~\ref{fig:steady} illustrates the growth of perturbations during the steady state of the heavy-jet model \modelid{U27}.
The top panels show cross-sections of $\log_{10}u_\infty$ at selected values of $z$, marked in the lower panels by vertical dotted white lines.
Inside the jet, $u_\infty \sim 100-200$ (dark red), and in the cocoon, $u_\infty \simeq u^r\amb \sim 0.1$ (dark blue).
The lower two panels show an azimuthally averaged map of the logarithm of  $u_\infty$ (middle panel) and its root-mean-square variations around the average value (lower panel). 
The jet boundary starts with a practically circular shape at $z=0$ and gradually becomes more and more distorted. 
This is visible in the upper panels of the figure, where the initial sinusoidal perturbation with $m=27$ is amplified throughout the collimation region, and intertwining `fingers' of jet and cocoon material appear. 
This structure is smeared above $z\sim 70$, probably because of the effect of the reflected shock that amplifies velocity variations through RMI.

\begin{figure*}
\includegraphics[width=1.0\textwidth]{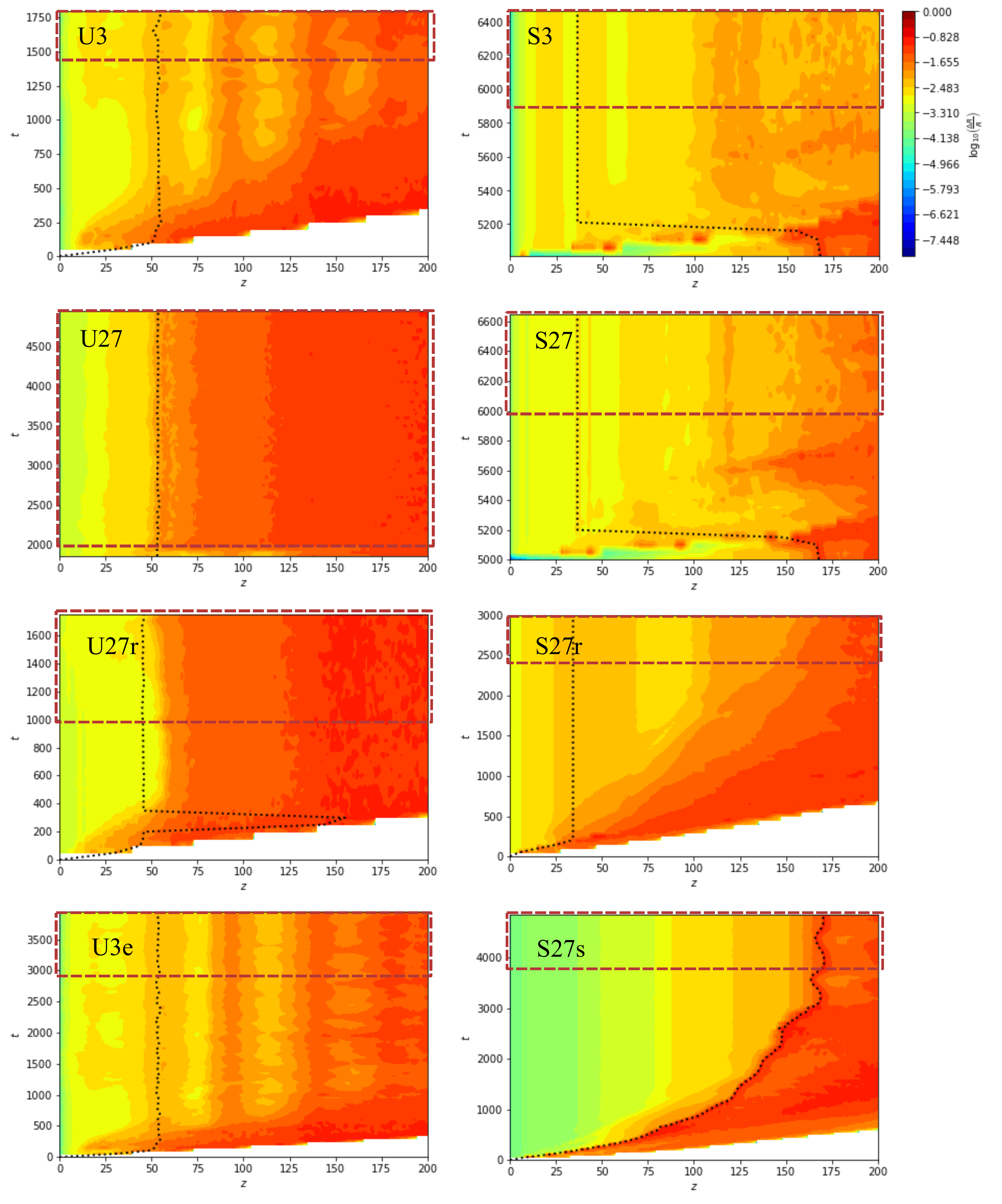}
 \caption{The growth of azimuthal dispersion of the contact discontinuity radius $\left(\Delta R/R\right)_{\rm d}$ as a function of time and $z$. Dotted line is the inferred position of the convergence point. Dashed rectangles mark the regions where the quantities were averaged in time to estimate the growth rate with $z$. 
 }\label{fig:mosaic}       
\end{figure*}

To quantify the amplitude of the perturbations, we first find the location of the contact discontinuity $R_{\rm d}(z, t, \varphi)$ by calculating the median value of $R$ for all the points with $1 < u^z < 5$. 
At each $z$ and $t$, this yields a range of radii, of which the maximal and the minimal values are shown in the two lower panels of Fig.~\ref{fig:steady} by white dotted lines.
Then we average $R_{\rm d}(z, t, \varphi)$ over the azimuthal coordinate $\varphi$ to get the instantaneous position of the discontinuity $R_{\rm d}(z, t)$. 
We measure the growth of the perturbations in the shape of the jet (corrugation of its surface) with the quantity $\Delta R(z, t)$, defined as the root-mean-square deviation of the jet radius $R_{\rm d}(z, t, \varphi)$ from the azimuthally averaged radius, $R_{\rm d}(z, t)$.
Fig.~\ref{fig:mosaic} shows how $\left(\Delta R / R\right)_{\rm d} = \Delta R(z,t)/R_{\rm d}(z,t)$ changes with time and altitude in different models. 
The dotted black line marks the location of the convergence point, $\zrec = \zrec(t)$. 

\begin{figure*}
\includegraphics[width=0.95\textwidth]{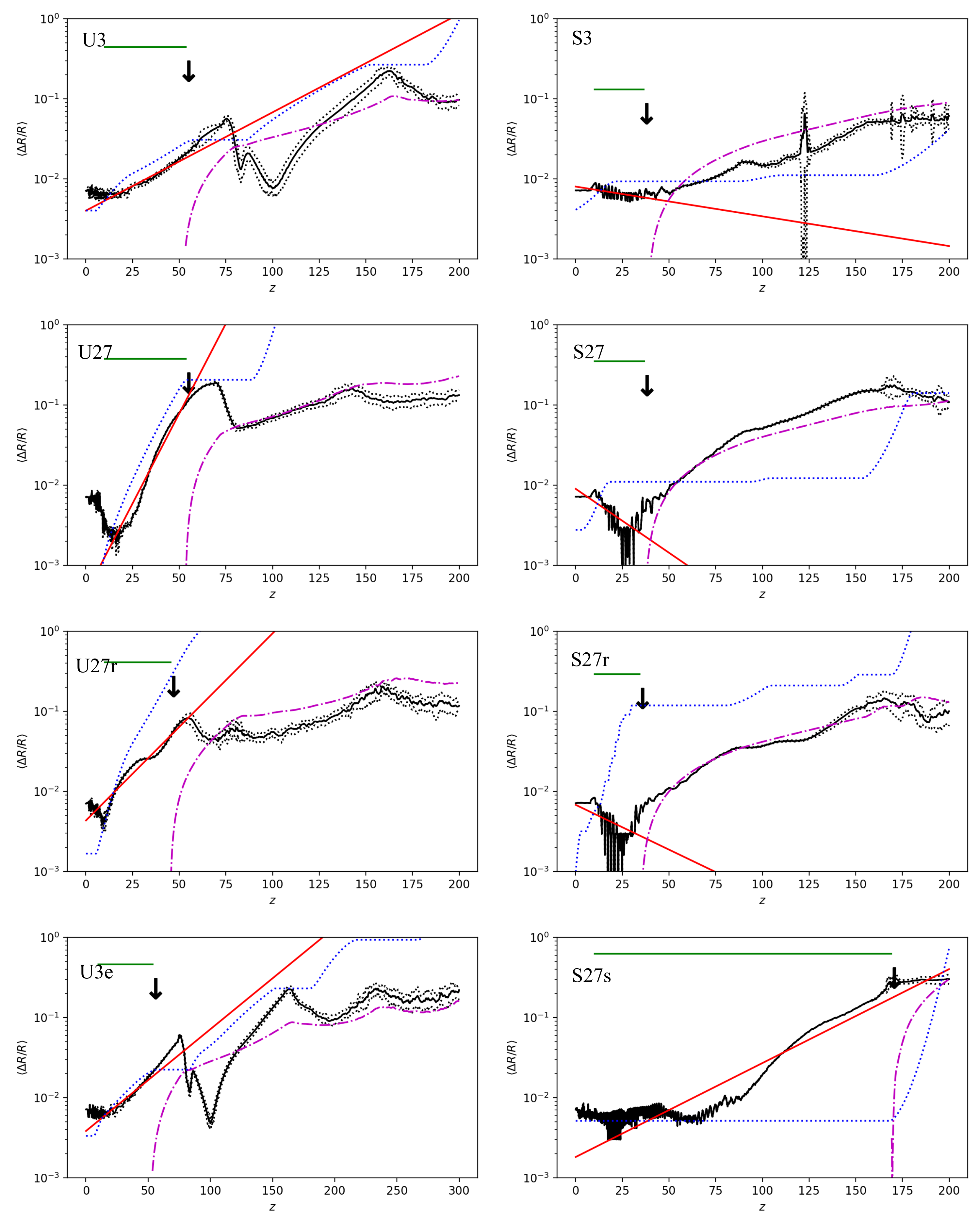}
 \caption{Time-averaged growth of the perturbations of the contact discontinuity. Solid black lines are the mean values of \ddr, dotted black lines show the variations with time (mean plus/minus root-mean-square deviation). 
 In each panel, the red solid line is an exponential fit to the black curve in the supposed linear-growth region ($10<z<\zrec$). 
 The blue dotted line is the growth expected from the linear RTI theory. 
Dot-dashed purple line is the approximate growth rate of the RMI (see Appendix~\ref{sec:app:RM}).
Position of the convergence point is shown by a vertical arrow. 
Green horizontal lines show the range of $z$ used to estimate the perturbation growth. 
 }
 \label{fig:slopes}
\end{figure*}

The quantity $\left(\Delta R / R\right)_{\rm d}(z,t)$ is averaged over time during the period of time we interpret as steady state (shown as red dashed boxes in Fig.~\ref{fig:mosaic}, see also Table~\ref{tab:res}), obtaining a time-averaged relative radius variation \ddr. 
In Fig.~\ref{fig:slopes}, we show \ddr\ as a function of $z$ for all the models studied in this work. 
In each panel we show with a dotted blue line the expected perturbation growth predicted by the linear theory (Eq.~\ref{E:RTI:dR}),
\begin{equation}
    \displaystyle \ddr_{\rm lin} (z) = \left(\frac{\Delta R}{R}\right)_0 {\rm e}^{\int_0^z \sqrt{-\A m \frac{1}{R} \frac{\diff^2R}{\diff z^2}}dz}.
\end{equation} 
The normalization for the integral was chosen to match the curves at $z = 10$. 
The Atwood numbers used for the analytic estimates in the individual models are given in Table~\ref{tab:mod}.
We define ${\cal A}$ globally using the flow parameters at injection ($z=0$).

The actual Atwood numbers on the jet-cocoon boundary may differ, but the introduced biases are small, because one relativistic inertia (of the jet or of the cocoon) always dominates over the other, reproducing values of $|{\cal A}| \simeq 1$.

We also estimate the mean growth rate $\eta = \frac{\diff }{\diff z}\ln\ddr$ in the range of altitudes from $10$ to $\zrec$. The
corresponding exponential fits are shown in Fig.~\ref{fig:slopes} with solid red lines. 
In Table~\ref{tab:res}, the values of $\eta$ measured in the simulation data are compared with the values predicted by the linear theory and averaged over the same range of $z$. 

In general, it can be seen that in the unstable, heavy-jet cases (left panels) there is a good agreement between the measured growth rates and the expectations from the linear theory. 
In the stable, light-jet models (right panels in Figs~\ref{fig:mosaic} and \ref{fig:slopes}), the growth rates are either consistent with zero within the fitting errors or show negative trends.
The only exception is the model \modelid{S27s}, where the structure of the transition region allows for an unstable layer with radially decreasing relativistic inertia. 
Above $\zrec$, all the models show positive growth of variations due to the RMI which operates on both stable and unstable models. 
The predicted growth of RMI is shown with purple dot-dashed lines and are calculated in Appendix ~\ref{sec:app:RM}.

When well-developed instabilities enter a non-linear regime, the discontinuity is smeared, and \ddr\ stops being a reliable criterion. 
In model \modelid{U27}, for example, this happens at $z\sim 70$ (see Fig.~\ref{fig:steady}). 
The smearing is evident in all the heavy-jet models (left panels in Fig.~\ref{fig:slopes}) as a drop in the value of \ddr\ above the convergence point. 
The reason for that is partly because of RTI entering a non-linear regime and partly due to the passage of the reflected collimation shock, which drives RMI and distorts the RTI `fingers' in the azimuthal direction. 
In models \modelid{U3} and \modelid{U3e}, where the instability growth rate is small, the perturbations are clearly affected by the passage of the shock at $z\sim 75$. 
However, the growth of RTI in these models is still visible up to $z\sim 180$ (note the large slope at $z\gtrsim 100$ inconsistent with the contribution of RMI).
In the light-jet models, the discontinuity shape criterion is reliable practically everywhere throughout the simulation domain.

In the light-jet models, which are expected to be stable against RTI, the primary factor of evolution is RMI, that turns on after the jet boundary is hit by the reflected collimation shock.
As we can see in Fig.~\ref{fig:slopes}, RMI predictions (purple dot-dashed lines) provide reasonably good fits to the perturbation growth in stable and sometimes in unstable models. 
We conclude that RMI is indeed the dominant instability above the convergence point, and its growth is limited only by non-linear effects. 

\subsection{Mixing and Bernoulli parameter}\label{sec:res:hg}

Without mixing, the jet material tends to conserve its Bernoulli parameter. 
This may be viewed as a consequence of mass and energy conservation laws, where the total mass flow rate $\dot{M} = \uppi R\jet^2 \rho u^z$, and jet power $L\jet = \uppi R\jet^2 \rho h \gamma u^z$ give the average Bernoulli parameter of $\langle \gamma_\infty\rangle = L\jet/\dot{M}$ (Eq.~\ref{eq:gamma_inf_avv}).
In an ideal case with no mixing, $\gamma_\infty$ tends to be constant along individual streamlines, and the distribution of mass or energy flux over $\gamma_\infty$ remains unchanged. 
The conservation of Bernoulli parameter and the clear difference between its values in the jet and in the cocoon allow to use it as a reliable criterion for mixing and mass entrainment. 

\begin{figure}
\includegraphics[width=1.0\columnwidth]{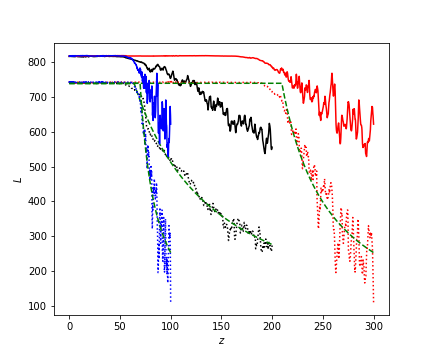}
 \caption{The total power coming in the flow with $u_\infty > 10$ (solid lines) and $u_\infty > 100$ (dotted lines) as a function of $z$ for models \modelid{U27} (black) and \modelid{U3e} (red). The blue lines show the model \modelid{U3e} compressed along the $z$ axis by a factor of 3. Dashed green lines are the fits to the dotted curves by a one-sided fractional function (Eq.~\ref{E:lofun}). 
 }\label{fig:hg:CD}       
\end{figure}

Fig.~\ref{fig:hg:CD} shows the jet power $L(z)$, estimated as the total power within streamlines with $u_\infty>10$ (solid lines) and $u_\infty>100$ (dotted lines) in models \modelid{U27} (black) and \modelid{U3e} (red). 
In our simulations the Bernoulli parameter can change only through mixing with the material from neighbouring streamlines, hence the figure shows the effect of mass loading caused by hydrodynamic instabilities.
In the heavy-jet models, strong mixing starts at a distance that scales $\propto \sqrt{m}$, consistent with the theoretical expectations for RTI. 
Apparently, mixing starts when the perturbations on the jet surface become non-linear and develop secondary instabilities.
The instabilities create a turbulent velocity field that shuffles streamlines with different Bernoulli parameters on the spatial scales down to the resolution limit of the simulations.
When mixing starts, the behaviour of the jet becomes similar to the regime considered in Section~\ref{sec:RTI:NL}.
The fact that energy loss starts in models \modelid{U3}, \modelid{U3e} only at about $z\sim 200$ means that the passage of the reflected collimation shock wave at $z\sim 70$, though it affects strongly the measured value of \ddr, does not instantaneously start mixing, but only distorts the jet boundary, making it difficult to trace its shape in the way done in Section~\ref{sec:res:inst}.

\begin{figure}
\includegraphics[width=1.0\columnwidth]{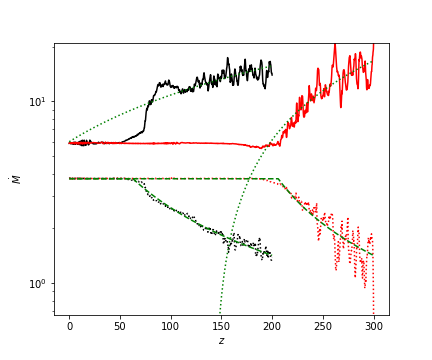}
 \caption{Total mass flow rate in the flow with $u_\infty > 10$ (solid lines) and $u_\infty > 100$ (dotted black and red lines) as a function of $z$ for models \modelid{U27} (black) and \modelid{U3e} (red).  Dashed green lines are fits to the dotted curves by a one-sided fractional function, dotted green lines are linear approximations for the solid curves downstream from $z_0$. 
 }\label{fig:mdetailed}       
\end{figure}

The curves shown in Fig.~\ref{fig:hg:CD} show the power loss from the jet core, defined by  different cut-off values of $u_\infty$.
They are well fitted by a one-sided linear fractional function 
\begin{equation}\label{E:lofun}
 \displaystyle   L(z) \simeq \frac{L_0}{1 + \max \left( \frac{z-z_{\rm min}}{\Delta z}, 0\right)},
\end{equation}
that may be viewed as a generalization of Eq.~(\ref{E:NL:ploss}).
Here, $z_{\rm min}$ and $\Delta z$ are free parameters which may be interpreted as the starting point of the non-linear phase and the characteristic mixing scale. 
For model \modelid{U27} (\modelid{U3e}), $z_{\rm min} = 64.1\pm 0.2$ ($209.5\pm 0.3$) and $\Delta z = 81.6\pm 0.3$ ($47.5\pm 0.3$).
Uncertainties here take into account only statistical errors of curve fitting, and thus are clearly underestimated.
The expected values of $z_{\rm min}$ correspond to the transition to the non-linear regime and are thus inversely proportional to $\eta \propto \sqrt{m}$ (Eq.~\ref{E:RTI:dR}).
The measured values are consistent with the predicted scaling within 10\%.
In Fig.~\ref{fig:hg:CD}, this is also illustrated by an overplotted re-scaled version of the power-loss curve for model \modelid{U3e}. 
As we have seen in Section~\ref{sec:RTI:NL}, the value of $z_{\rm mix}$ has a clear physical meaning related to jet mixing but independent of the properties of the initial perturbations. Fig.~\ref{fig:mdetailed} shows the corresponding mass flow of jet material with $u_\infty>10$ (solid) and $u_\infty>100$ (dotted) in models \modelid{U27} (black) and \modelid{U3e} (red). 
Evidently, the relativistic core ($u_\infty > 100$) looses mass as well as energy (because of the overall slow-down), while the mildly relativistic sheath gains mass through entrainment. 
The rapid changes in the mass flow start at roughly the same $z$ coordinate as the jet deceleration. 
The mass-flow grows roughly linearly with $z$, as expected in a cylindrical jet (Eq. \ref{eq:dot_M_lin}), though strong variations are present corresponding to large-scale non-linear structures within the jet.

\subsection{Effects of rotation }\label{sec:disc:KHI}

\begin{figure*}
\includegraphics[width=0.9\textwidth]{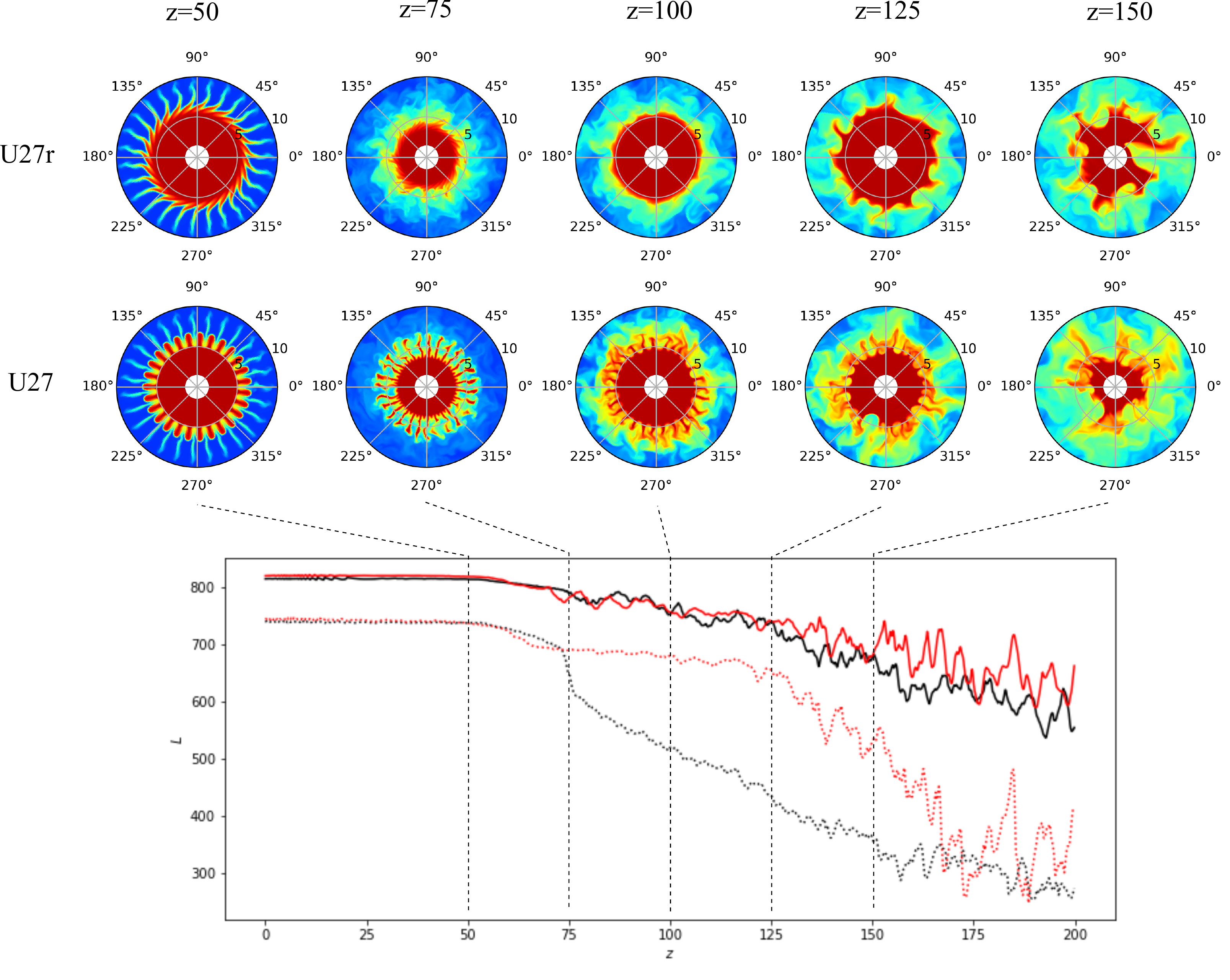}
 \caption{Effects of rotation on RTI development and mixing in heavy-jet models. Lower panel: the same as Fig.~\ref{fig:hg:CD} for models \modelid{U27} (black) and \modelid{U27r} (red). Upper panels show $z=$const cross-sections of the two models. Colour-coded is $u_\infty$ with the limits of $0.1$ and $300$, the scale is logarithmically uniform. 
 }\label{fig:hg:rotation}       
\end{figure*}

As one can see from Table~\ref{tab:res}, adding slow rotation ($\Omega \simeq 0.1$) to the initial velocity of the jet does not significantly change the growth rate in the linear regime. 
There are, however, differences in the non-linear regime, as can be seen in Fig.~\ref{fig:hg:rotation}. 
For the core of the jet ($u_\infty > 100$), the power loss curve behaves practically the same as in the non-rotating case. 
However, there is a strong difference in the mixing at the jet periphery ($u_\infty > 10$).
A probable reason is that the angular frequency near the jet boundary is comparable to the RTI growth rate, and thus affects the initial penetration depth of the RTI `fingers' (see the upper panels of Fig.~\ref{fig:hg:rotation}). 
However, mixing slows down the rotation of the jet, and the power loss close to $z\sim 200$ is practically identical with and without rotation. 

We do not see any signatures for Kelvin-Helmholtz instability (KHI) in the rotating-jet models, which is a probable outcome of the non-linear interaction of the velocity shears in $\varphi$ and $z$ directions. 
The role of KHI in the stability of relativistic jets was considered in a number of papers (see, e.g., \citealt{hardee06} and references therein).
While the surface of a relativistic supersonic jet is stable to KHI \citep{bodo04}, the jet as a whole may be a subject to global instabilities. 
The growth rate of such a mode may be estimated as
\begin{equation}
    \sigma_{\rm KHI} \simeq \frac{\gamma\jet\sqrt{\eta}}{1+\gamma\jet^2\eta} k_z,
\end{equation}
where $\eta = h\jet \rho\jet / h\amb \rho\amb$, and $k_z$ is the wavenumber along the $z$ axis. 
In our models, both stable and unstable, $\sigma_{\rm KHI} \sim 0.3k_z$. 
For a wave numbers $k_z$ of the order several (corresponding to several times the thicknesses of the transition layer), the growth rate is considerably larger than the RTI growth rate. 
The most likely reason for the absence of growing Kelvin-Helmholtz modes is our boundary conditions at the jet base. 
The velocity components at the jet base are fixed, excluding all unstable modes that have considerable phase velocity along the jet. 

\section{Observational implications}\label{sec:disc}

Our study shows that once a hydrodynamic jet enters the non-linear mixing stage, at about several collimation distances it quickly loses its momentum and slows down, heating the ambient medium. 
It is interesting to check if there is observational evidence supporting these findings.

\subsection{Application to AGN}\label{sec:disc:AGN}

Typical Lorentz factors in AGNs are about several to several tens \citep{hovatta09}. 
For some well-studied galactic nuclei (such as \object{M87} and \object{3C84}) the shape of the jet is known directly from observations. 
A typical FR-I radiogalaxy jet has a parabolic inner part that may be viewed as the collimation region, which becomes conical outside a certain radius \citep{hada19, kovalev20}.
In \object{M87} the transition to a conical shape roughly coincides with the \object{HST-1} feature, which is located $\sim100$\pc\ away from the BH and is often associated with the convergence point of a collimation shock \citep{Stawarz06}.
The location of \object{HST-1} also coincides with the estimated radius of the BH sphere of influence, inside which the BH gravitational potential dominates over that of the surrounding stars (e.g. \citealt{ferrarese}).
The sphere of influence radius may be estimated as $r_{\rm i} \simeq GM_{\rm BH}/\sigma_{\rm v}^2$, where $\sigma_{\rm v}$ is the velocity dispersion of the stars populating the galactic nucleus.
If the surrounding gas is in virial equilibrium with the stars, its speed of sound is of the same order as $\sigma_{\rm v}$, and $r_{\rm i}$ also has the physical meaning of the Bondi radius.

The density profile inside $r_{\rm i}$ is expected to be shallow, with a slope of $-\frac{\diff\ln\rho\amb}{\diff \ln z} \sim 1-1.5$ (\citealt{yuan_review} and references therein), followed by a steeper profile outside.
In particular, for \object{M87}, observational data suggest $\rho\amb \propto z^{-1}$ inside the Bondi radius \citep{russell15, park_RM}.
It is also confirmed by global numerical simulations of accretion and ejection flows in galactic centres reported in \citet{guo22}.
This kind of a shallow slope may extend beyond the sphere of influence within a cocoon, as hot cocoons formed by relativistic jets tend to smear the background density gradients \citep{Harrison18}.
For a conical jet propagating through isothermal medium with a density profile $\rho\amb \propto z^{-1}$, the mass flow rate behaves as $\dot{M} \propto z$ (Eq.~\ref{eq:dMdot_dz}). 
Accordingly the jet Lorentz factor should decrease as $\gamma \sim <\gamma_\infty> \propto z^{-1}$ (Eq. \ref{eq:gamma_inf_avv}), consistent with the observed slow-down of the jet of M87 beyond the \object{HST-1} knot reported by \citet{park_gamma}. 

In cases where the density slopes differ significantly inside and outside the sphere of influence, the position of the convergence point relative to $r_{\rm i}$ may be crucial for jet survival.
For high-power jets, collimation is relatively weak, and $z_*$ is expected to be located well beyond $r_{\rm i}$.
As a result, mass entrainment due to interface instabilities is suppressed, and the jets retain high Lorentz factors all the way to their heads, consistent with the observational properties of FR-II radiogalaxies.
For example, in \object{Cygnus~A}, the jet power is $\sim4\times10^{45}\ergl$ \citep{Ito08}, $\sim40$ times higher than the jet of \object{M87} \citep[e.g.][]{Bicknell96}, whereas the BH mass is about half of the BH mass in \object{M87}. 
Assuming similar pressure and density within the cores of the two galaxies, and noting that $\zrec$ scales with $\sqrt{L}$ (Eq.~\ref{E:disc:zrec}), while $r_{\rm i}$ scales linearly with the BH mass, we get that $\zrec\simeq10 r_{\rm i}$. 
For \object{Cygnus A}, jet collimation takes place mostly in the steep density region where collimation is less efficient. 
As a result two effects take place: (i) the linear growth rate of the RTI (Eq.~\ref{E:RTI:dR}) decreases, that pushes the transition towards the non-linear regime further downstream along the jet; (ii) the mass load rate during the non-linear stage decreases as a result of decreasing ambient mass density.
For a conical jet in an isothermal gas with a density profile $\rho\amb \propto z^{-2}$, the mass load and jet deceleration scale logarithmically with $z$. 
The combination of these two factors allows the jets in objects like \object{Cygnus~A} to remain relativistic over much longer distance than the jet of \object{M87}.
As stronger jets are decelerated less efficiently, one also might expect a general dichotomy in jet properties.
The difference between FR-I and FR-II jets may therefore be determined by the location of the convergence point inside or outside the BH sphere of influence.

\subsection{Application to X-ray binaries}

The measured Lorentz factors in microquasar jets are typically $2-5$ or even smaller \citep{saikia19}, while the observed opening angles are in most cases  $\lesssim10^\circ$ \citep{miller06}. 
For a population of freely expanding jets, these values are at odds, as one would expect jet opening angles of order $1/\gamma$. 
This tension can be reconciled by two possibilities:
(i) the actual Lorentz factors in the jets are higher than the values implied by the observations \citep[e.g.][]{miller06}, or 
(ii) the jets are collimated by the pressure of a cocoon. 
To estimate whether the jets are collimated we can look at the quantity \citep{matzner03, bromberg11}
\begin{equation}\label{eq:Ltilde}
\tilde{L}=\frac{L_j}{\Sigma_{\rm h}\rho\amb c^2},
\end{equation}
where $\Sigma_{\rm h}$ is the cross-sectional area of the jet head and $\rho\amb$ is the ambient medium rest-mass density. 
$\tilde{L}$ affects the propagation velocity of the jet head, $\beta_{\rm h}$, and the pressure in the cocoon. 
When $\tilde{L}\gg1$, the jet internal energy density is much larger than the energy density in the ambient medium, and $\beta_{\rm h}\simeq 1$. 
The energy that flows from the head into the cocoon is reduced by a factor $1-\beta_{\rm h}$, resulting in a low cocoon pressure, which is unable to efficiently collimate the jet. 
When $\tilde{L}\ll1$, the head moves at a non-relativistic velocity $\beta_{\rm h} \simeq\tilde{L}^{1/2}$ \citep{matzner03}, and most of the jet power flows into the cocoon leading to a high cocoon pressure and efficient jet collimation. 
\citet{bromberg11} showed that, as long as $\tilde{L}<\theta_0^{-4/3}$, where $\theta_0$ is the jet injection angle, the cocoon pressure is high enough to collimate the jet.

An uncollimated jet maintains a constant opening angle resulting in $\Sigma_{\rm h}=\uppi\theta_0^2 z_{\rm h}^2$, where $z_{\rm  h}$ is the location of the jet head. 
If the jet propagates in a medium with a uniform density, $\tilde{L}$ decreases with distance, and we can find a critical distance $z_{\rm h,\, c}$, for which, if $z_{\rm h}>z_{\rm h,c}$, the jet becomes collimated. 
Estimating the jet power by the Eddington luminosity \citep{fender04}, 
\begin{equation}
    L_{\rm Edd} = \frac{4\uppi GMc}{\varkappa_{\rm T}} \simeq 10^{39} \frac{M}{10\Msun} \ergl,
\end{equation}
where $\varkappa_{\rm T} \simeq 0.35~{\rm cm^2\, g^{-1}}$ is the Thomson scattering cross-section for solar metallicity, and assuming an ambient medium number density of $n\amb \sim 10^{-3}\cmc$ \citep{heinz02}, we get 
\begin{equation}
\begin{array}{lcc}
z_{\rm h,\,c} & = &\displaystyle\left(\frac{L\jet}{\uppi\rho\amb c^3\theta_0^{2/3}}\right)^{1/2}\\
& \simeq & \displaystyle 0.08 \sqrt{\frac{L\jet}{10^{39}\ergl}} \sqrt{\frac{10^{-3}\cmc}{n\amb}} \left(\frac{10\deg}{\theta_0}\right)^{1/3} \pc.\\
\end{array}
\end{equation}
As typically microquasar jets extend to distances of order parsecs \citep{corbel11} from their sources, it follows that the jets must be collimated. 

To estimate the collimation pressure and calculate the location of the collimation shock convergence point, we use the expression for the cocoon pressure from \citet[][eq. 16]{bromberg11} for the case of a relativistic head
\begin{equation}\label{eq:Pc}
P\amb  \simeq \displaystyle\left(\frac{L\jet^2\rho\amb^3\theta_0^2}{(z_{\rm h}/c)^4}\right)^{1/5}.
\end{equation}
Substituting this in Eq.~(\ref{E:disc:zrec}), we get a collimation distance of 
\begin{equation}
\begin{array}{l}
    \displaystyle\zrec  \simeq \left(\frac{L_j}{\rho\amb c^3}\right)^{3/10}\left(\frac{z_{\rm h}^2}{\theta_0}\right)^{1/5}\\
    \displaystyle \qquad{} \simeq  0.3 \left(\frac{z_{\rm h}}{1\pc}\right)^{2/5}\left(\frac{L\jet}{10^{39}\ergl}\right)^{3/10}\\
   \displaystyle  \qquad{} \qquad{} \qquad{} \left( \frac{\theta}{10\deg}\right)^{-1/5}\left(\frac{n\amb}{10^{-3}\cmc}\right)^{-3/10}\pc ,\\
\end{array}
\end{equation}
implying that RTI and RMI have enough time to grow to non-linear scales and lead to entrainment of heavy material into the jet. 
This can limit the Lorentz factor of the jet material and provide a natural explanation for the low values observed. 
We thus conclude that microquasar jets are collimated and are likely loaded with baryons. 
The degree of mixing and its effect on the Lorentz factor need to be investigated quantitatively elsewhere.  

\subsection{Gamma-ray bursts}

Within the framework of a hydrodynamic jet, the observed GRB emission is generated in internal shocks formed when jet elements accelerated to different Lorentz factors collide \citep{Rees94,sari_piran}. 
The variations in Lorentz factors originate from variations in Bernoulli parameter, which can be traced back either to the injection mechanism, or to mass loading taking place along the jet. 
The latter process is naturally associated with RTI and RMI in the non-linear stage forming turbulence that mixes ambient matter into the jet. 
In such a model all the physical properties of the jet become highly variable in space and time, suggesting intrinsic variability of the GRB emission.
Hence, variability of GRBs is an important test for the interface instabilities in the jet. 

GRBs consist of two classes: long GRBs (lGRBs) with typical durations $\gtrsim 2\,$s, originating from collapsing massive stars \citep{piran_review, Meszaros06}, and short GRBs (sGRBs) with durations $\lesssim 2\,$s, associated with merging compact objects, likely binary neutron stars \citep{Nakar07,Berger14}. 
Beside the difference in duration, comparison between the temporal structure of sGRB lightcurves with the first $2\,$s of lGRB lightcurves shows similar behaviour
\citep{McBreen01, Nakar02, Bhat12}. 
As lGRB ligtcurves do not show signs of evolution during the prompt phase \citep[e.g.][]{Yoshida17}, identical mechanism should be responsible for the variability in both GRB types.
If the variability originates from perturbations amplified by interface instabilities, similar variability patterns suggest similar amount of collimation and mass entrainment in the confining medium.

To evaluate the effect of mixing on the jets of both GRB types, we calculate the location of $z_*$ relative to the radius of the confining medium $\ra$ at the time of jet breakout. 
Following \citet{bromberg11}, we evaluate the cocoon pressure as the energy density in the cocoon at the time of breakout, approximating the cocoon shape as a cylinder 
\begin{equation}\label{E:GRB:pamb}
P\amb \simeq\frac{L\jet t_{\rm b}(1-\hbeta)}{\uppi\theta_{\rm c}^2 \ra^3},
\end{equation}
where $\theta_{\rm c} \simeq R\amb/\ra$ is the cocoon opening angle, where $R\amb$ is the cocoon cylindrical radius, $\hbeta$ is time-averaged jet head velocity, and $t_{\rm b} = \ra / c \hbeta$ is the breakout time of the jet.
Substituting Eq.~(\ref{E:GRB:pamb}) in Eq.~(\ref{E:disc:zrec}), we obtain
\begin{equation}
\frac{z_*}{\ra}\simeq \left(\frac{\hbeta}{1-\hbeta}\right)^{1/2}\theta_{\rm c}.
\end{equation}
For lGRBs, $\theta_{\rm c}$ is similar to the opening angle of the jet material after it exits the star \citep{bromberg11, Harrison18}, constrained to be $\sim0.1$ by observations \citep[e.g.][]{Fong12, Goldstein16}, and $\hbeta\sim0.3$ \citep[e.g.][]{bromberg12, Mizuta13, Harrison18}, leading to $z_*=0.06\ra$. 
In sGRBs, $\theta_{\rm c}$ and $\hbeta$ are constrained by numerical simulations to be $\sim0.5$ and $\sim0.7$, respectively \citep{bromberg18, Gottlieb21}, thus $z_*\simeq0.8\ra$. 
These estimates show that in the case of lGRBs there is enough time for RTI and RMI to evolve to non-linear scales and mix heavy stellar material into the jets.
In sGRBs, on the other hand, the shock convergence point is located close to the edge of the dynamical ejecta at the time of breakout, thus interface instabilities cannot grow much beyond the linear scale, and the resultant jet is expected to be less variable than in the case of lGRBs.
These predictions are consistent with results of hydrodynamic simulations that studied the properties of post-breakout long and short GRB jets \citep{Gottlieb19,Gottlieb21}.
However, they are inconsistent with the observational data showing similar variability in both types of GRBs. 
Therefore, we suggest that the observed variability is unrelated to interface instabilities.
Instead, its origin may be related to intermittency of the central machine, which in both cases is likely a compact hyper-accreting accretion disc around a newly formed black hole. 

\section{Discussion and conclusions}\label{sec:conc}

Our simulations confirm that hydrodynamic relativistic jets collimated by external dense media are subject to Rayleigh-Taylor and Richtmyer-Meshkov instabilities that grow on the contact discontinuity between the jet and its cocoon.
Rayleigh-Taylor instability (RTI) develops in the collimation region on the length scale close to the collimation length scale. 
Richtmyer-Meshkov instability (RMI) becomes important after the collimation shock reflects from the jets axis and reaches the contact discontinuity. 
The estimated growth rates of both instabilities agree well with the analytic predictions. 
We show that, for appropriate perturbation modes, both instabilities reach a non-linear stage at a distance of the order collimation length.

In the non-linear regime, we expect the jet to mix efficiently with its environment.
The interaction includes mass loading into the jet and energy leaks, leading to a rapid decrease in the Lorentz factor of the jet material.
This non-linear mixing regime allows for an analytic solution well consistent with our simulations in the case of a uniform-density ambient medium. 
The presence of a steep density gradient, like in a case when a jet is breaking out of a dense core, slows down the mixing and mass loading.

In real astrophysical objects, the amount of mass loading in relativistic jets is defined by the interplay between collimation processes that trigger interface instabilities and the pre-existing background gradients in density and pressure.
In particular, the rapid decline in Lorentz factor observed in \object{M87} is well explained by interaction with an ambient medium with a density profile $\rho \propto z^{-1}$, consistent with the observational estimates. 

In gamma-ray bursts (GRBs),
interface instabilities can operate 
efficiently in the region between the convergence point and the edge of the confining medium (stellar envelopes in long GRBs and dynamical mass ejecta in short GRBs). 
We show that the location of the convergence point relative to the edge is vastly different in the two populations and that the instabilities can lead to variable emission predominantly in long GRBs. 
This prediction is not supported by observations, which show similar variability patterns in both GRB types. We suggest that the observed variability likely originates from the modulations in the jet injection process.
Numerical studies of intermittent hydrodynamic lGRB jets by \citet{Gottlieb20} show that they cannot produce bright $\gamma$-ray emission due to extensive mixing with the stellar envelope. The existence of weak magnetic fields was shown to reduce mass loading, leaving pockets of jet material with high values of Bernoulli parameter that can accelerate to relativistic velocities and efficiently radiate the GRB emission \citep{Gottlieb20b, Gottlieb21}.
Thus, the fact we observe jet emission in lGRBs is indicative for the possible presence of magnetic fields stabilizing the jets . 

While weak magnetic field with $B^2/4\pi\rho \sim 0.01-0.1$ was shown to stabilize relativistic jets, stronger magnetization leads to current-driven instabilities such as pinch \citep{Chatterjee2019} or kink \citep{BT16}. 
The features produced by pinch instability sometimes resemble RTI fingers, but their development conditions and spatial structures differ.
Note that the structures visible in the simulations of \citet{Chatterjee2019} are two-dimensional and in our setup would require time-dependent boundary conditions.

Our calculations show that microquasar jets are likely collimated by the ISM and that the convergence point is located far enough from the jet head to allow for the interface instabilities to become non-linear and mix ISM material into the jet. This can resolve the apparent discrepancy between the low Lorentz factors determined from superluminal features in the jets and the small jet opening angles measured.

Our results suggest a new explanation to the observed dichotomy between FR-I and FR-II radiogalaxy jets. 
The dichotomy was explained in the past by the growth of Kelvin-Helmholtz instability on the jet boundary \citep[e.g.][]{Kaiser97,Perucho05,Meliani09,Perucho10}, kink instability in magnetic jets \citep{Tchekhovskoy16} or by mass entrainment from stellar winds \citep[e.g.][]{Komissarov94, Perucho14, Wykes15}.
We suggest that the observed differences may also be related to mass entrainment induced by RTI and RMI.
In this scenario, in low-power FR-I jets the collimation shock converges on the jet axis inside the galactic core (more precisely, within the BH sphere of influence), which has a relatively flat mass distribution profile, allowing for RTI and RMI to become non-linear and mix heavy material into the jet. 
In high-power FR-II jets, the shock convergence point is located well outside the core, in regions with steep density gradients, resulting in a slower perturbation growth during the linear stage and lower mixing rate during the non-linear stage allowing the jets to retain their relativistic properties and survive over long distances.

\section*{Acknowledgements}
We would like to thank E. Nakar and A. Levinson for helpful discussions and the anonymous referee for useful suggestions. The study was supported by an ISF grant 1995/21 (PA) and by an ISF grant 1657/18 (OB).

\section*{Data Availability}

The data underlying this article will be shared on reasonable request to the corresponding author.



\bibliographystyle{mnras}
\bibliography{mybib} 




\appendix

\section{Non-linear Rayleigh-Taylor regime in an expanding jet}\label{sec:app:fermi}

To show how different modes are amplified in the case of an expanding jet, where collimation plays the role of gravity, let us consider an axisymmetric jet of the known shape $R_{\rm d} = R_{\rm d}(z)$ and fixed cocoon velocity $\beta\ll 1$. 
For simplicity, we will also assume a constant Atwood number $\A$. 

For  given azimuthal mode $m$, $\displaystyle k = \frac{m}{R}$, and the instability increment is 
\begin{equation}
    \sigma = \sqrt{\A g k } =  \sqrt{\frac{\A g}{R}m}.
\end{equation}
As in Section~\ref{sec:res:inst}, effective gravity depends on the shape of the contact surface
\begin{equation}
    g \simeq \beta^2 \frac{\diff^2R_{\rm d}}{\diff z^2}.
\end{equation}
In the linear regime, each mode grows independently, and creates radial deformation
\begin{equation}\label{E:app:dR}
    \displaystyle \Delta R(z)= \Delta R(z_0) e^{\sqrt{\A m }\int_{z_0}^z \sqrt{-\frac{1}{R_{\rm d}}\frac{\diff^2R_{\rm d}}{\diff z^2}} dz}.
\end{equation}
Growth of the particular mode is limited either by the intrinsic thickness of the transition layer, or by interaction with itself (deformation of the surface becomes comparable to the wavelength of the relevant mode). 

\begin{figure}
\includegraphics[width=\columnwidth]{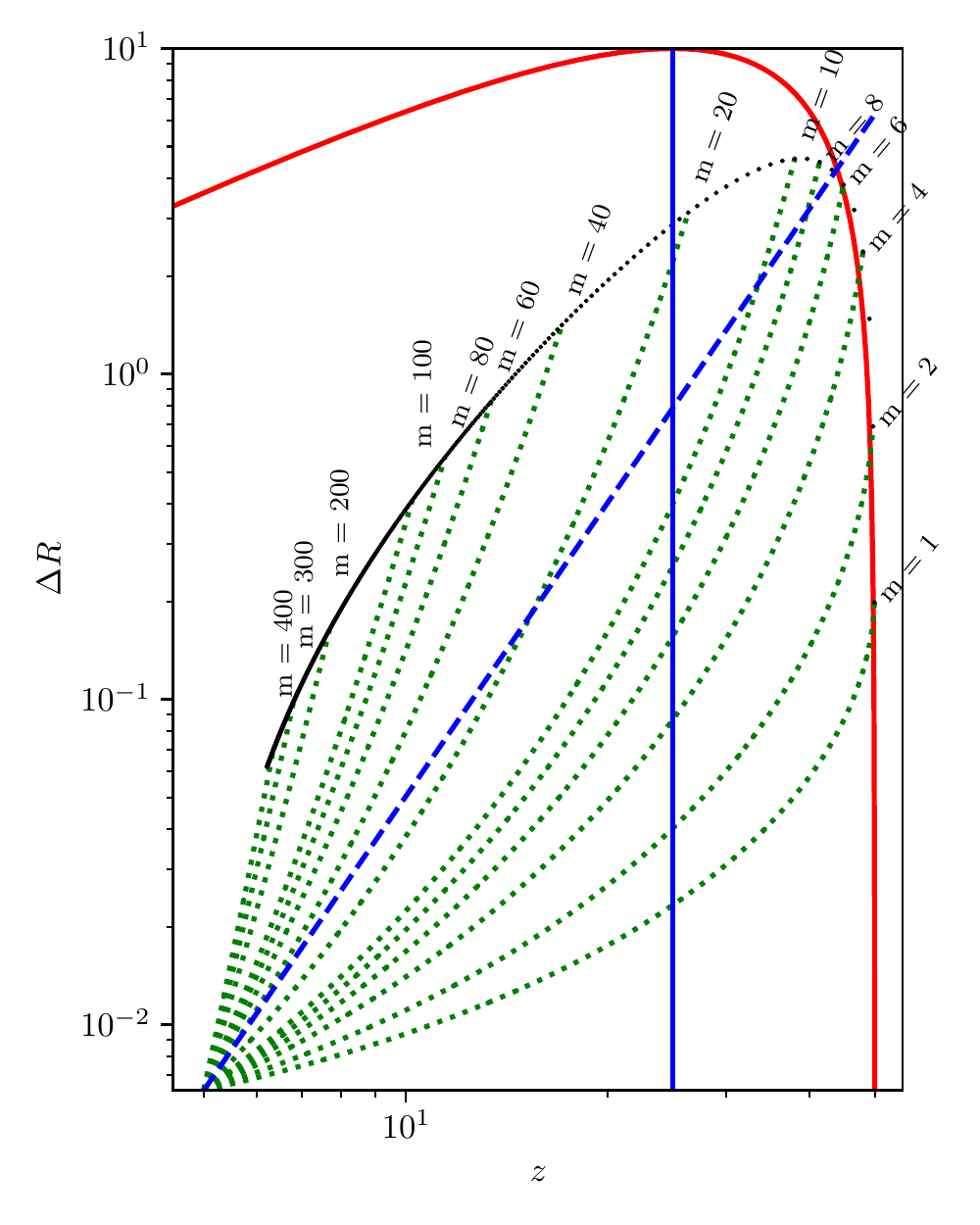}
 \caption{ Growth and saturation of different RTI modes in a parabolic jet. 
 Green dotted lines are the growing perturbations, and the black dots show the transition of the particular mode to the non-linear regime.
 The red solid line is the jet boundary, blue vertical line corresponds to $z = \zrec/2$. 
 The blue dashed line is the self-similar scaling $\Delta R \propto z^3$. 
 }\label{fig:parmplot}       
\end{figure}

As the instability grows faster for shorter wavelengths, the limit would be encountered first for higher-order modes. 
Condition for the transition towards the non-linear regime may be set as $\displaystyle\Delta R = \frac{2\uppi R_{\rm d}}{m}$, that implies, after substituting Eq.~(\ref{E:app:dR}), 
\begin{equation}\label{E:nonlinear:ln}
    \ln \left(m\delta \frac{R_{\rm d}(z_0)}{R_{\rm d}(z)}\right) + \sqrt{ \A m }\int_{z_0}^z \sqrt{\frac{1}{R_{\rm d}}\frac{d^2R_{\rm d}}{dz^2}} dz = 0,
\end{equation}
where $\displaystyle\delta = \frac{\Delta R(z_0)}{2\uppi R_0}$. 
For given $m$ and $R_{\rm d} = R_{\rm d}(z)$, this may be solved for $z$, yielding the distance at which particular mode $m$ saturates.
In Fig.~\ref{fig:parmplot}, we show how individual modes grow and saturate for a parabolic jet shape $\displaystyle R_{\rm d}(z) = 4R_{\rm max} \frac{z}{\zrec} \left( 1- \frac{z}{\zrec}\right)$. 
Initial relative perturbation of $\delta = 10^{-4}$ was set at $z_0 = 1$, $R_{\rm max} = 5$, $\zrec = 50$. 

Parabolic jet expansion is a good approximation for the shape of the collimation shock or for the contact discontinuity upstream from the maximal radius.
Then acceleration decreases, and the expansion of the jet is effectively frozen (see Section~\ref{sec:res:jet}). 
To estimate the strongest perturbation mode in a jet that has already experienced collimation, it seems reasonable to choose the dominating mode somewhere near the maximal expansion in the parabolic model ($z= \zrec/2$). 
Substituting the parabolic jet shape in Eq.~(\ref{E:nonlinear:ln}) yields
\begin{equation}\label{E:app:arcsin}
    \ln \frac{m \delta}{\zeta (1-\zeta)} + 2\sqrt{2\A m} \left. \arcsin \sqrt{1-\zeta}\right|^{\zeta}_{\zeta_0} = 0,
\end{equation}
where $\zeta = z/\zrec$, $\zeta_0 = z_0 / \zrec$. 
Substituting $\zeta = 1/2$ and $\zeta_0 =0$ allows to write an equation for $m$
\begin{equation}
    \ln m \delta + \frac{\uppi}{2} \sqrt{2 \A m}  = 0.
\end{equation}
Approximately, the solution is
\begin{equation}
    m = \frac{2 \ln^2(m\delta)}{\uppi^2\A} \simeq \frac{2 \ln^2\delta }{\uppi^2\A}.
\end{equation}
For example, $\delta = 10^{-4}$ and $\A\sim 1$ yields $m \simeq 10$. 
The stronger the initial perturbation, the smaller the dominating mode number $m$. 

Before the maximal expansion point, in the $\zeta \ll 1$ limit, Eq.~(\ref{E:app:arcsin}) may be approximately re-written as
\begin{equation}
    \ln \frac{m\delta }{\zeta} + \sqrt{2\A m} (\zeta_0 - \zeta) \simeq 0.
\end{equation}
For $\zeta \gg \zeta_0$, this implies a scaling $m \propto \zeta^{-2}$, and $\Delta R \propto R/m \propto \zeta^3$ (one extra power of $\zeta$ comes from the radius).
This regime is analogous to the Fermi -- von Neumann asymptotics for the classic RTI problem \citep{Fermi53}.

\section{Richtmyer-Meshkov instability}\label{sec:app:RM}

Development of this instability type requires a shock crossing the contact discontinuity. 
In our simulations, this happens downstream from the convergence point, when the reflected collimation shock reaches the boundary of the jet. 

Following \citet{richtmyer54}, the perturbation growth in Richtmyer-Meshkov instability may be viewed as a result of rapid velocity perturbation growth during the passage of the shock itself, followed by a period of approximately linear growth of $\Delta R = \Delta v \tau +$const. 
Here, $\tau$ is the time in the cocoon co-moving frame. 
As in the RTI problem, we will ignore all the gradients in $z$ direction.

The value of $\Delta v$ may be estimated by integrating the instantaneous perturbation growth triggered by the rapidly changing acceleration field near the shock front (see \citealt[equation 3]{richtmyer54})
\begin{equation}
    \frac{\diff \Delta v}{\diff \tau} = \A k g(\tau) \Delta R,
\end{equation}
where $g(\tau) = \diff u^r / \diff \tau$ is locally measured inertial force, that is assumed to vary much faster than all the other quantities on the right-hand side of the equation.
This implies an expression for the velocity perturbation after the collision with the shock 
\begin{equation}
    \Delta v \simeq \A m \frac{\Delta R}{R} \Delta \beta^r,
\end{equation}
where $\Delta \beta^r$ is the velocity jump in the shock.
After the shock, velocity perturbations cause linear growth in the surface shape perturbations $\Delta R$
\begin{equation}
    \Delta R = \Delta R_0 \left(1+\A m \frac{\Delta \beta^r}{\beta} \frac{\Delta z}{R_0}\right).
\end{equation}
In Fig.~\ref{fig:slopes}, we show the approximate contribution of RMI to the growth of $\Delta R / R$, assuming $\Delta \beta^r = \beta$ and normalized for $\Delta R$ at $z=\zrec$. 
Though there is no physical reason to assume the two velocities equal, if the shock is strong and stationary, $\Delta \beta^r \sim \beta \simeq 1$. 
For most of the models, and especially for the light-jet models, RMI gives a good fit for the perturbation growth after the convergence point.

\section{The structure of the jet-sheath transition}\label{sec:app:boundary}

Here, we resolve the structure of the jet boundary and make very specific assumptions about its structure. 
This is partly done to avoid numerical noise from sharp jumps in hydrodynamical parameters, and partly to ensure the jet is stable to the instability modes with the lengths comparable to the spatial resolution. 
Making the boundary smooth allows to control the particular perturbation modes that are amplified by the linear regime of RTI. 

The boundary of the jet is located at $R_{\rm j} = 5$ and has a characteristic width of $\Delta R_{\rm j} = 0.5$. 
There are multiple ways to set the initial structure of the boundary. 
The way the physical parameters change between the jet and the sheath is in fact extremely important for the stability properties of the flow. 
To match the variables on both sides of the boundary, we use a sigmoid function of the form
\begin{equation}\label{E:app:smoothfun}
	s(x) = \frac{1}{2} \left( 1+ \tanh x\right).
\end{equation}
This is a monotonic function changing from 0 to 1 while its argument changes from $-\infty$ to $+\infty$, and already very close to its limits (about 0.12 and 0.88, respectively) at $x = \pm 1$. 
We will say that the physical quantity $q = q(R)$ follows a `simple step' if 
\begin{equation}
q(R) = q_{\rm j} + \left(q_{\rm sh}- q_{\rm j}\right) s\left( \frac{R-R_{\rm j}}{\Delta R_{\rm j}}\right).
\end{equation}
If we assume that density and $u^z$ follow simple steps through the boundary between a relativistic jet and non-relativistic ambient medium, and the jet is under-dense with respect to the ambient medium, then relativistic inertia $A=\gamma^2h\rho$ becomes a non-monotonic function of $R$. 
Indeed, in the non-relativistic regime $A \simeq \rho$, and for $\gamma^2 \gg 1$ and $P\gtrsim \rho$, $A\simeq \gamma^2 P$. 
Thus, $A$ is an increasing function of radius on the outer (non-relativistic) side, and decreases with radius inside the jet. 
Existence of a minimum in $A$ creates a zone unstable to short-wavelength perturbations (see Appendix~\ref{sec:app:smoothRT}), that may be seen in simulation \modelid{S27s}. 
This result is still valid for the convective stability criterion $\ppardir{R}{h^2A} >0$ we derive in Appendix~\ref{sec:app:smoothRT}.

Instead of setting a simple step in primitive variables, in all the simulations except \modelid{S27s} we set simple steps in $A$ and in relativistic Bernoulli number
\begin{equation}
\gamma_{\infty} = h\gamma.
\end{equation}
The third independent parameter, pressure, was assumed constant. 
For given $A$ and $\gamma_\infty$, density and velocity components may be found analytically. 
In particular, $\gamma$ is expressed as a function of $A$, $\gamma_\infty$, and $P$ as
\begin{equation}
 \gamma = \frac{A}{8P\gamma_\infty} \left( \sqrt{1+ \frac{16 P \gamma_\infty^2}{A}}-1\right).
\end{equation}
Density is  found as
\begin{equation}
\rho = \frac{A}{\gamma_\infty\gamma}.
\end{equation}
To find the velocity components, we assume that physical velocity component ratios $u^r / u^z$ and $u^\varphi/u^z$ also follow simple steps throughout the transition region. 

\section{Linear Rayleigh-Taylor stability for short wavelengths}\label{sec:app:smoothRT}

Local Rayleigh-Taylor stability of the stratified transition region between the jet and the sheath may be considered in terms of local constant-gravity space-time.
We will be using Rindler space-time \citep[section 6.6]{MTW}, Kottler-M\o{}ller coordinates \citep{munoz_rindler}
\begin{equation}
    ds^2 = - (1+gy)^2 dt^2 + dx^2+dy^2+dz^2,
\end{equation}
where $g$ is effective gravity. 
The only non-zero Christoffel symbols are 
\begin{equation}
    \Gamma_{ytt} = g(1+gy),
\end{equation}
\begin{equation}
    \Gamma_{tyt} = \Gamma_{tty} = - g(1+gy).
\end{equation}
Without the loss of generality, we will consider the region near $y=0$, where $g_{tt}=-(1+gy)^2\simeq -1$.
First, consider continuity equation
\begin{equation}\label{E:GR:con}
    \left( \rho u^k \right)_{; k} = 0.
\end{equation}
In coordinates, this may be written as
\begin{equation}\label{E:GR:con:coord}
    \ppardir{k}{(1+gy) \rho u^k} = 0.
\end{equation}
Stress-energy equation in its general form is
\begin{equation}\label{E:GR:EST}
    \left( h\rho u^k u_i + P \delta_i^k\right)_{;k} = 0.
\end{equation}
The set up of the problem has three Killing vectors ($t$, $x$, and $z$), that allows in all the components of the stress-energy equation except $y$ to replace covariant derivatives with partial.
Expanding Eq.~(\ref{E:GR:EST}) and taking into account continuity equation (Eq.~\ref{E:GR:con}),
\begin{equation}\label{E:GR:general}
    \frac{1}{h} u^k \ppardir{k}{h u_i} = - g (1+gy) (u^t)^2 \delta_i^y   - \frac{1}{h\rho} \pardir{i}{P}.
\end{equation}
Let us assume that the unperturbed solution is in hydrostatic balance in $y$ direction. Then,
\begin{equation}\label{E:GR:static}
    \pardir{y}{P} = -g (1+gy)(u^t)^2 h\rho.
\end{equation}
Let us consider linear perturbations of the velocity $u^i = U^i + \delta u^i$, density $\rho+\delta \rho$, pressure $P+\delta P$, and enthalpy $h+\delta h$.
Linearizing Eq.~(\ref{E:GR:general}) and assuming all the perturbations scale as $\delta \propto {\rm e}^{\i q_k x^k}$ yields
\begin{equation}
    \i \omega \delta u_i + \frac{1}{h}\delta u^y \ppardir{y}{h U_i} = 
    - g(1+gy) \delta (u^t)^2 \delta_i^y + \frac{\delta(h\rho)}{(h\rho)^2} \pardir{i}{P},
\end{equation}
where $\omega = q_k U^k$. 
Here, we assumed that all the unperturbed quantities depend only on $y$, and $U^x = U^y = 0$. 
Unperturbed quantities depend on $y$ only.
We also neglected pressure perturbation $\delta P$, assuming dynamic equilibrium approximately holds on the time scales considered. 
Perturbation of $\delta (u^t)^2$:
\begin{equation}
    \delta (u^t)^2 = 2U^t \delta u^t = - \frac{2}{(1+gy)^2} U^t \delta u_t .
\end{equation}
The spacial components of this equation acquire the form
\begin{eqnarray}
\i \omega \delta u_t + \frac{1}{h}\delta u^y \ppardir{y}{h U_t} =0 \label{E:GR:dut} \\
\i \omega \delta u_x = 0\\
\i \omega \delta u_y = \frac{2g}{1+ay}  U^t \delta u_t + \frac{\delta (h\rho)}{(h\rho)^2} \pardir{y}{P} \label{E:GR:duy} \\
\i \omega \delta u_z + \frac{1}{h}\delta u^y \ppardir{y}{h U_z} = 0 .\label{E:GR:duz} 
\end{eqnarray}
Here, we also neglected the pressure variation term in $\delta (h\rho)$, and $U^t$ is denoted as $\gamma$. 
The first two equations are redundant, and the remaining system is solvable if supplemented with the entropy conservation equation and an expression for $\delta (h\rho)$. 
In general,
\begin{equation}\label{E:GR:hrho}
    \delta (h\rho) = \delta \rho + \frac{\Gamma}{\Gamma-1}\delta P.
\end{equation}
If we neglect Eulerian variations of $\delta P$, $\delta (h\rho) = \delta \rho$.

Entropy conservation its general form is written as
\begin{equation}
    u^k \ppardir{k}{\ln \left( \frac{P}{\rho^\Gamma}\right)} = 0.
\end{equation}
Perturbation of the adiabatic equation is
\begin{equation}
    \delta u^y \ppardir{y}{\ln \left( \frac{P}{\rho^\Gamma}\right)} + \i \omega \left( \frac{\delta P}{P} - \Gamma \frac{\delta \rho}{\rho}\right)=0.
\end{equation}
If pressure perturbation is zero, the above equation implies
\begin{equation}\label{E:GR:ad}
    \frac{\delta \rho}{\rho} = \frac{1}{\i \Gamma \omega} \delta u^y \ppardir{y}{\ln \left( \frac{P}{\rho^\Gamma}\right)}.
\end{equation}
Substituting (\ref{E:GR:dut}) to (\ref{E:GR:duy}) yields
\begin{equation}
\delta u^y \left( 1 + \frac{2 g }{h\omega^2} U^t \ppardir{y}{h U_t }\right) = - \frac{1}{\i \omega} \pardir{y}{P} \delta \frac{1}{h\rho}.
\end{equation}
Then, substituting $\delta(h\rho)$ using (\ref{E:GR:hrho}) and (\ref{E:GR:ad}) yields
\begin{equation}
    \delta u^y \left( 1 - \frac{2g}{h\omega^2}  U^t \ppardir{y}{h U_t}\right) = \frac{1}{\Gamma \omega^2} \delta u^y \frac{1}{h^2\rho} \pardir{y}{P}\ppardir{y}{\ln \frac{P}{\rho^\Gamma}}.
\end{equation}
Then, let us introduce physical Lorentz factor $\gamma = \sqrt{- U_t U^t}$, such as $U^t = \sqrt{-g^{tt}}\gamma$. Then, we can use Eq.~(\ref{E:GR:static}) to replace the vertical pressure gradient $\pardir{y}{P} = -\gamma^2 g h\rho / (1+gy)$. 
Near $y=0$, the dispersion equation may be then simplified as
\begin{equation}
    \omega^2 = \frac{g U_t U^t}{1+gy} \left( -\frac{1}{\Gamma h} \ppardir{y}{\ln \frac{P}{\rho^\Gamma}} + 2 \ppardir{y}{\ln h |U_t|}\right),
\end{equation}
or 
\begin{equation}
    \omega^2 \simeq  - g\gamma^2 \left[ -\frac{1}{\Gamma h} \ppardir{y}{\ln \frac{P}{\rho^\Gamma}} + 2 \frac{\uppartial}{\uppartial y}\ln \left(h \gamma\right)\right].
\end{equation}
Inside the transition layer, $P$ changes much less than density, that allows to rewrite the dispersion relation as $\omega^2 \simeq -\gamma^2 g \frac{\uppartial}{\uppartial y}\ln\left(\gamma^2 h^3\rho\right)$. 

\bsp	
\label{lastpage}
\end{document}